\documentclass[%
 reprint, 
 superscriptaddress,
nofootinbib,
 amsmath,amssymb,
 aps,
 pra,
]{revtex4-2}

\usepackage[export]{adjustbox}
\usepackage{graphicx}
\usepackage{dcolumn}
\usepackage{bm}
\usepackage[colorlinks=true,citecolor=blue,linkcolor=blue, urlcolor=blue]{hyperref}

\usepackage{color}

\usepackage[acronym,shortcuts]{glossaries}

\usepackage{lipsum}
\usepackage{braket}

\newcommand{\todo}[1]{}
\renewcommand{\todo}[1]{{\color{red} TODO: {#1}}}
\newcommand{\question}[1]{}
\renewcommand{\question}[1]{{\color{red} QUESTION: {#1}}}
\newcommand{\note}[1]{}
\renewcommand{\note}[1]{{\color{red} NOTE: {#1}}}

\renewcommand\Re{\operatorname{Re}}

\DeclareMathOperator{\Tr}{Tr}

\newacronym{ZNE}{ZNE}{zero-noise extrapolation}
\newacronym{ARMA}{ARMA}{autoregressive moving average}
\newacronym{SchWARMA}{SchWARMA}{Schr\"{o}dinger wave autoregressive moving average}
\newacronym{NISQ}{NISQ}{noisy intermediate-scale quantum}
\newacronym{QNS}{QNS}{quantum noise spectroscopy}
\newacronym{RB}{RB}{randomized benchmarking}
\newacronym{CPMG}{CPMG}{Carr-Purcell-Meiboom-Gill}

\begin{document}

\glsdisablehyper
\preprint{APS/123-QED}

\title{Analyzing the impact of time-correlated noise on zero-noise extrapolation}

\author{Kevin Schultz}
\email{kevin.schultz@jhuapl.edu}
\affiliation{Johns Hopkins University Applied Physics Laboratory, Laurel, MD, 20723, USA}

\author{Ryan LaRose}
\affiliation{Department of Computational Mathematics, Science, and Engineering, Michigan State University, East Lansing, MI, 48823, USA}
\affiliation{Unitary Fund}

\author{Andrea Mari}
\affiliation{Unitary Fund}

\author{Gregory Quiroz}
\affiliation{Johns Hopkins University Applied Physics Laboratory, Laurel, MD, 20723, USA}

\author{Nathan Shammah}
\affiliation{Unitary Fund}

\author{B. David Clader}
\thanks{Current Affiliation: Goldman, Sachs \& Co, New York, NY}
\affiliation{Johns Hopkins University Applied Physics Laboratory, Laurel, MD, 20723, USA}

\author{William J. Zeng}
\affiliation{Unitary Fund}
\affiliation{Goldman, Sachs \& Co, New York, NY}

\date{\today}

\begin{abstract}
Zero-noise extrapolation is a quantum error mitigation technique that has typically been studied under the ideal approximation that the noise acting on a quantum device is not time-correlated. In this work, we investigate the feasibility and performance of zero-noise extrapolation in the presence of time-correlated noise. We show that, in contrast to white noise, time-correlated noise is harder to mitigate via zero-noise extrapolation because it is difficult to scale the noise level without also modifying its spectral distribution. This limitation is particularly strong if ``local'' gate-level methods are applied for noise scaling. However, we find that ``global'' noise scaling methods, e.g., global unitary folding, can be sufficiently reliable even in the presence of time-correlated noise. We also introduce gate Trotterization as a new noise scaling technique that may be of independent interest.
\end{abstract}

\maketitle



\section{Introduction}

The theory of fault-tolerant error-corrected quantum computation may result in speed-ups in a number of computations, most notably the factoring of numbers using Shor's algorithm \cite{shor1994algorithms,chuang1995quantum,shor1999polynomial}, but also in quantum simulation and chemistry \cite{cirac2012goals,jones2012faster,houck2012chip,georgescu2013quantum,o2016scalable}, linear systems \cite{harrow2009quantum,clader2013preconditioned,pan2014experimental,wossnig2018quantum}, and other areas \cite{shor2002introduction,childs2010quantum,venegas2012quantum,schuld2015introduction,montanaro2016quantum,biamonte2017quantum}.  While ongoing progress has improved the performance of  individual qubits and has allowed quantum computers to scale to larger number of qubits, current systems are not sufficiently performant for useful fault-tolerant operations.  Despite this apparent limitation, we are in or rapidly nearing a regime where quantum systems could perform useful computations without (or with less) error correction, the so called \ac{NISQ} era \cite{preskill2018quantum}.

In this regime of \ac{NISQ} computations, it is imperative that any potential errors be reduced or mitigated in order to maximize the utility from these imperfect devices and/or small distance codes.  A number of potential techniques have been proposed to mitigate errors in the \ac{NISQ} regime including quantum control \cite{viola1999dynamical,viola2003robust,brixner2004quantum,vandersypen2005nmr,khaneja2005optimal,d2007introduction,biercuk2009optimized,dong2010quantum,souza2012robust,machnes2018tunable,lucarelli2018quantum,bukov2018reinforcement}, decoherence-free subspaces \cite{lidar1998decoherence,bacon1999robustness,kwiat2000experimental}, readout error mitigation \cite{Chow2010,Chow12,Maciejewski2021,Bravyi2021}, Pauli frame randomization \cite{kern2005quantum,wallman2016noise,ware2021experimental}, and optimal compilation \cite{venturelli2018compiling,zulehner2018efficient,dueck2018optimization,venturelli2019quantum,davis2020towards,tan2020optimal}.  One recently-proposed technique motivated by \ac{NISQ} limitations is \ac{ZNE} \cite{temme2017error,Li_2017_PRX,kandala2019nature,giurgica2020digital,He_2020_PRA,kim_scalable_2021}. This aims to mitigate the impacts of any errors on a computation by performing a series of computations with \textit{scaled} error levels then post-processing to interpolate to the \textit{zero-noise limit} of the computation.

\ac{ZNE} techniques have been primarily investigated under the assumption that the errors to be mitigated are uncorrelated in time. On the other hand, time-correlated noise (in particular $1/f^{\alpha}$ noise) has been widely observed in physical systems including
superconducting devices \cite{bylander2011:fnoise, yan2013:fnoise,burnett2014:fnoise,muller2015:fnoise,burnett2019:fnoise}, quantum dots \cite{basset2014:fnoise,chan2018:qns}, and spin qubits \cite{struck2020:fnoise}. In NISQ devices, such as those offered by the IBM Quantum Experience, evidence of correlated noise has been observed both indirectly through the use of dynamical error suppression~\cite{pokharel2018demonstration,niu2022pulse} and directly through \ac{QNS} estimation of the noise~\cite{murphy2021universal}. This has been further substantiated by recent studies that have suggested the dynamics of such devices are more accurately captured by non-Markovian models~\cite{tripathi2021suppression,zhang2021predicting}. 

To estimate the noise present in these real physical systems, one can use \ac{QNS} \cite{alvarez2011:qns,szakowski2017:qns,pazsilva2017:qns} wherein the outcomes of a set  of distinct control pulses or circuits are analyzed.  Key to this approach is that while these different probe sequences may in fact represent identical circuits under ideal conditions, they interact with any noise present in different ways.  This can be understood through the filter function formalism \cite{cywinski2008:fff,PhysRevLett.113.250501} which describes the ``frequency response'' of a given probe sequence.  Broadly speaking, the impacts of noise (in terms of fidelity) are approximately proportional to the integral of the product of the power spectrum of the noise with the filter function of the control, called an overlap integral.  In what follows, we will show how this intuition can also be applied to different \ac{ZNE} schemes in the presence of temporally correlated dephasing noise.

The recently developed \cite{schultz2020schwarma} and experimentally validated \cite{murphy2021universal} \ac{SchWARMA} technique provides a natural mechanism for the exploration of so-called \textit{digital} \ac{ZNE} techniques~\cite{dumitrescu2018cloud,giurgica2020digital,He_2020_PRA} that operate at the gate level in a quantum circuit.  Building on techniques from classical time-series modeling in statistics and signal processing, \ac{SchWARMA} was conceived as a highly flexible mechanism for simulating a wide-range of spatiotemporally correlated errors in quantum circuits. 

In the following, we first review the \ac{SchWARMA} modeling and simulation formalism as well as a concise overview of \ac{ZNE} and discuss different methods for scaling noise.  Next, we show how these different schemes are impacted by time-correlated dephasing noise despite the fact that they behave equivalently for uncorrelated noise.  We then interpret these noise scaling schemes using the language of filter functions and show that these results are well described by the intuition provided by the filter functions. Our findings indicate that, for time-correlated noise, the noise scaling method known as {\it global unitary folding} \cite{giurgica2020digital, larose2020mitiq} produces more accurate noise-scaled expectation values and \ac{ZNE} results.

\section{Background}

\subsection{Time-correlated noise: The SchWARMA model}

Consider a single-qubit Hamiltonian
\begin{equation}\label{eq:dep_ham}
    H(t)=H_z(t)+H_c(t)
\end{equation}
consisting of a semiclassical dephasing noise component $H_z(t)$ along with a deterministic idealized control component $H_c(t)$ corresponding, for example, to the external driving induced by laser pulses.  If we further define 
$H_z(t)=\eta(t)\sigma^z$ with $\eta(t)$ a wide-sense stationary Gaussian stochastic process, we can say that this noise process is not time-correlated if $\mathbb E[\eta(t)\eta(t')]=\mathbb E[\eta(|t-t'|)\eta(0)]=0$ for all $t\neq t'$, where $\mathbb E(\cdot)$ represents the average over many statistical realizations. $\sigma^i$, $i=x,y,z$ are the Pauli matrices. Equivalently, we can say that the noise process is time-correlated if the power spectrum 
\begin{equation} \label{eq:spectrum}
    S_\eta(\omega)=\int_0^{\infty} dt\,\mathbb E[\eta(t)\eta(0)]e^{-i\omega t}
\end{equation}
is not constant as a function of $\omega$ (i.e., not a ``white'' process).
This semiclassical noise setting is the standard setting for \ac{QNS} \cite{alvarez2011:qns,szakowski2017:qns,pazsilva2017:qns} and is an alternative to general open quantum systems approaches that consider couplings to quantum baths. The semiclassical noise approximation assumes that the bath is in thermal equilibrium and at infinite temperature, yielding regimes with no back action on the environment from the qubits, as well as equal populations of qubit states after long term decay \cite{kubo1963stochastic,haken1973exactly,vcapek1993haken,cheng2004stochastic,gardinerhandbook,cheng2005microscopic,van1992stochastic}.

\begin{figure}[t!]
    \includegraphics[width=0.9\columnwidth]{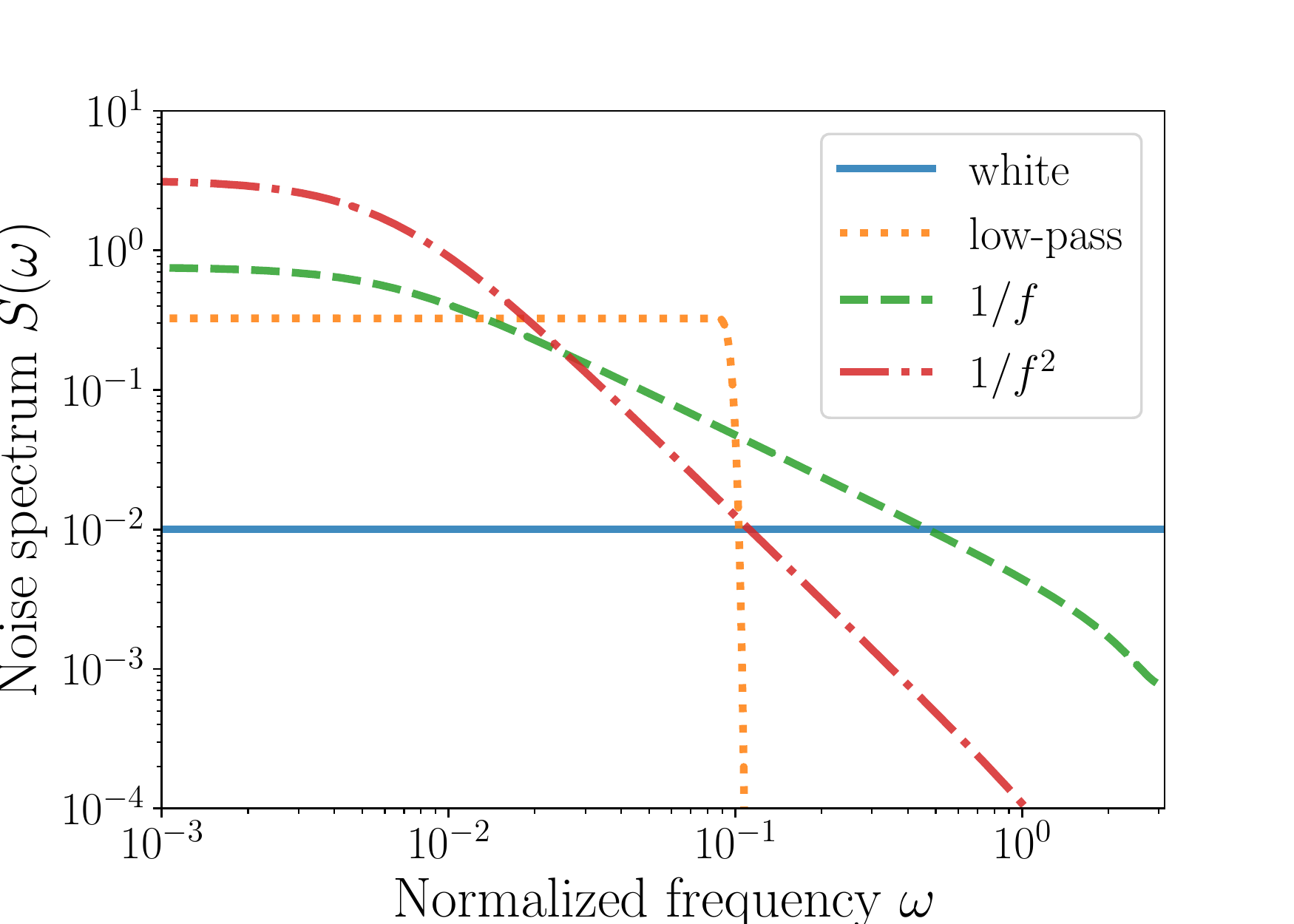}
    \caption{Noise power spectrum of four different dephasing SchWARMA noise models corresponding to white noise, low-pass noise, $1/f$ noise and $1/f^2$ noise. These noise models are used in Sec.~\ref{sec:results} to test the effect of time-correlated noise on zero-noise extrapolation.}
    \label{fig:noise_spectra}
\end{figure}

In the \ac{SchWARMA} modeling approach \cite{schultz2020schwarma}, the impact of the continuous time Hamiltonian in \eqref{eq:dep_ham} is modeled in a quantum circuit formalism by inserting correlated $Z$-error operators after each ``gate'' determined by the control $H_c$. This is accomplished by generating a time-correlated sequence of rotation angles $y_k$ defined from independent Gaussian inputs $x_k$ using an \ac{ARMA} model \cite{whittle1963prediction,box2015time},
\begin{equation}\label{eq:ARMA}
   y_k = \underbrace{\sum_{i=1}^p a_{i}
    y_{k-i}}_{AR}+\underbrace{\sum_{j=0}^{q} b_{j}x_{k-j}}_{MA}\,,
\end{equation}
where the set $\{a_i\}$ defines the autoregressive portion of the model, and $\{b_j\}$ the moving average portion with $p$ and $q+1$ elements of each set respectively. 
The time correlations are defined via the resulting power spectrum 
\begin{equation} \label{eq:spectrum_schwarma}
    S_y(\omega)=\frac{\left|\sum_{k=0}^q  b_k \exp(-ik\omega)\right|^2}{\left|1+\sum_{k=1}^pa_k\exp(-ik\omega)\right|^2}\,,
\end{equation}
and ARMA models can approximate \textit{any} discrete-time power spectrum to arbitrary accuracy \cite{holan2010arma}. For the scope of this work we focus on the four paradigmatic noise spectra shown in Fig.~\ref{fig:noise_spectra}, namely: white noise, low-pass noise, $1/f$ noise and $1/f^2$ noise.

Dividing the circuit trajectory defined by $H_c(t)$ into consecutive gates $G_k$, the \ac{SchWARMA} approach models the impact of correlated noise $H_z(t)$ by adding in a random $Z(\theta_k)=\exp(iy_k\sigma^z)$ after each gate, which can then be Monte Carlo averaged to produce an expectation value.
This model can be extended to multi-qubit Hamiltonians  
\begin{equation}
    H(t) = \sum_{j=1}^n\eta_j(t)\sigma^z_j+H_c(t)\,,
\end{equation}
by generating independent, yet identically defined, \ac{SchWARMA}-generated errors on each qubit.  In principle, these could of course be heterogeneous and correlated between qubits.

\subsection{Zero-noise extrapolation with colored noise}

Zero-noise extrapolation (ZNE) is an error mitigation technique which relies on the ability to increase the noise in a quantum circuit~\cite{temme2017error,Li_2017_PRX,Endo_2018_PRX}. Like other error mitigation techniques, the target is to estimate an expectation value
\begin{equation}
    E(\lambda) := \Tr [ \rho(\lambda) O]
\end{equation}
at zero noise. The noise scale factor $\lambda$ dictates how much the base noise level $\lambda = 1$ is scaled in the quantum circuit which prepares the system density matrix $\rho$, and $O$ is a problem-dependent observable. The key insight of ZNE is to (i) evaluate $E(\lambda)$ at several noise scale factors $\lambda \ge 1$, then (ii) fit a statistical model to the collected data and infer the zero-noise value $E(\lambda \rightarrow 0)$. We refer to these two steps as noise scaling and inference, respectively.

Compared to other error mitigation techniques, zero-noise extrapolation requires very few additional quantum resources. Correspondingly, it has received some attention in recent literature; e.g., it was implemented in Refs.~\cite{kandala2019nature,dumitrescu2018cloud,giurgica2020digital,larose2020mitiq,Lowe_2021_PRR,Mari_2021_PRA} and in~\cite{kim_scalable_2021} on twenty six superconducting qubits to produce results competitive with classical approximation techniques. References~\cite{dumitrescu2018cloud,giurgica2020digital,He_2020_PRA} formally introduced digital noise scaling, in which noise is scaled at a gate-level  without pulse-level control. 

While ZNE is straightforward to implement and requires relatively few additional quantum resources, the quality of the solution depends critically on both the inference and noise-scaling method and can be improved by a correct characterization of the hardware noise. In this work, we fix the inference method by assuming a particular noise model and focus on the effects of the noise-scaling method.





\subsubsection{Noise scaling methods}

\begin{figure}
    \centering
    \includegraphics[width=\columnwidth]{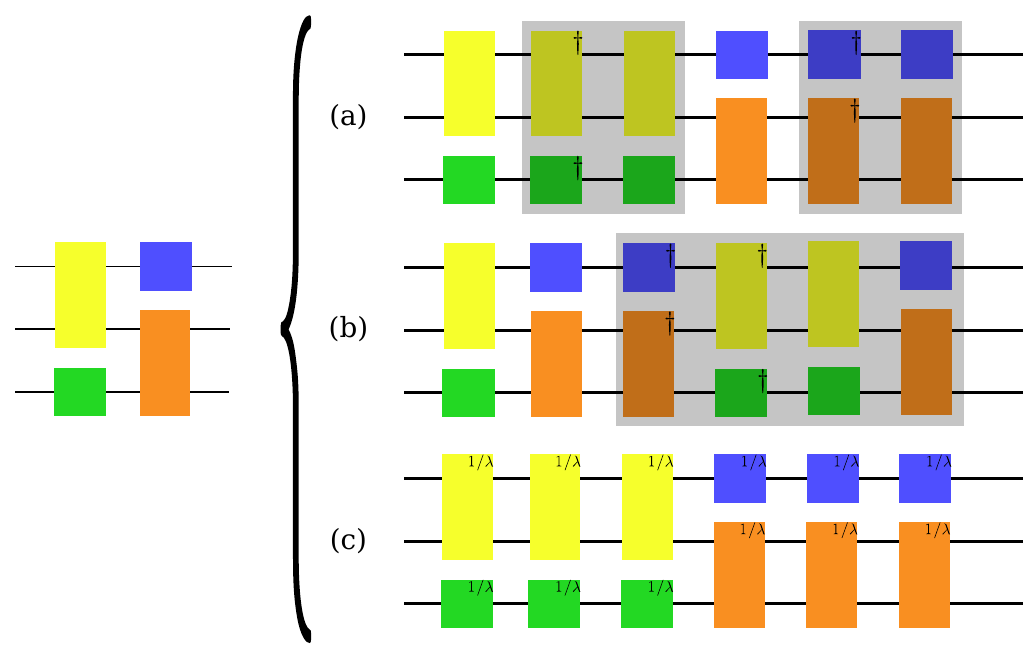}
    \caption{A sample three-qubit circuit with four gates under the action of three digital noise scaling methods we consider in this work. (a) Local folding, in which each gate $G$ gets mapped to $G \mapsto G \left( G^\dagger G \right)^n$ for scale factor $\lambda = 2n - 1$. (b) Global folding, in which the entire circuit $C$ gets mapped to $C \mapsto C \left( C^\dagger C \right)^n$. In (a) and (b), grey shading shows the ``virtual gates'' which logically compile to identity. (c) Gate Trotterization, in which $G \mapsto \left( G^{1 / \lambda} \right) ^ \lambda$ for each gate $G$.}
    \label{fig:scaling}
\end{figure}

\paragraph{Ideal noise scaling}

In a purely theoretical setting, the ideal way of scaling the noise would be to multiply the Hamiltonian $H_z$ in Eq.\ \eqref{eq:dep_ham} by a constant $\sqrt{\lambda}$:
\begin{equation}\label{eq:ideal_scaling}
    H'(t)=\sqrt{\lambda}H_z(t)+H_c(t).
\end{equation}
Equivalently, the scale factor can be absorbed into a redefinition of the stochastic noise amplitude: $\eta'(t) = \sqrt{\lambda} \eta(t)$. From  Eq.\ \eqref{eq:spectrum_schwarma}, it is evident that the noise power spectrum gets scaled by $\lambda$,
\begin{equation} \label{eq:spectrum_scaled}
    S_{\eta'}(\omega)= S_{\sqrt{\lambda} \eta}(\omega) = \lambda\, S_\eta (\omega).
\end{equation}

If one could directly control the noise, this would be the ideal way of scaling its power and, therefore, the ideal way of applying zero-noise extrapolation. 
In simulations using \ac{SchWARMA}, noise can be scaled by transforming the numerator coefficients $b_k\to\sqrt{\lambda}b_k$. 
In a typical experimental scenario, of course, one cannot directly control the noise of a quantum device. Even in instances where it is possible to scale the noise spectrum through e.g., manipulating the master clock \cite{murphy2021universal} or flux lines, precise characterization of the native noise spectrum and calibration of the noise injection would be required.
Due to these difficulties in directly scaling noise, several indirect noise scaling techniques have been proposed and applied in recent literature. We define several of these in the following subsections (see Fig.~\ref{fig:scaling} for an overview) in order to analyze their performance in the presence of time-correlated noise in Sec.~\ref{sec:results}.

\paragraph{Pulse stretching}
The intent of pulse stretching is to scale the impacts of the noise on the system by ``stretching'' the underlying control Hamiltonian, replacing \eqref{eq:dep_ham} with
\begin{equation}
    H(t) = H_z(t)+\frac{1}{\lambda}H_c(t/\lambda)\,,
\end{equation}
for some dimensionless time-scaling factor $\lambda$.  In principle, this scales the impacts of the noise by increasing the overall time duration of the circuit. More precisely, if we define  $t'=t/\lambda$, the density operator $\rho(t')$ of the system evolves with respect to the effective Hamiltonian:
\begin{equation}\label{eq:pulse_stretching}
    H'(t') = \lambda\,  H_z(\lambda t') + H_c(t')\,.
\end{equation}
The corresponding noise power spectrum is:
\begin{align} \label{eq:spectrum_stretched}
    S_{\eta'}(\omega)& = \lambda ^2 \int_{0}^{\infty} dt'\,\mathbb E[\eta(\lambda t')\eta(0)]e^{-i\omega t'} \nonumber\\
    &= \lambda \int_{0}^{\infty} dt\,\mathbb E[\eta(t)\eta(0)]e^{-i\omega t/\lambda} = \lambda\, S_\eta(\omega/\lambda).
\end{align}
From the equation above, it is evident that for a white (constant) spectrum, pulse stretching can be used to effectively scale the noise power by $\lambda$ as in the ideal case defined in Eq.\ \eqref{eq:spectrum_scaled}. In fact, the equivalence between the ideal noise scaling and the pulse-stretching technique was already shown in Ref.\ \cite{temme2017error}, under the hypothesis of a quantum state $\rho$ evolving according to a master equation with a time-independent noise operator acting as $\mathcal L (\rho)$ (more details about the consistency between our findings with the results of Ref.\ \cite{temme2017error} are given in Appendix \ref{app:consistency}). 
On the other hand, Eq.\ \eqref{eq:spectrum_stretched} shows that, for a colored spectrum, pulse-stretching does not exactly reproduce the ideal noise scaling defined in \eqref{eq:spectrum_scaled}. Indeed, on the r.h.s.\ of  Eq.\ \eqref{eq:spectrum_stretched} we observe that the original spectrum is also stretched with respect to the frequency variable $\omega$. This fact is a manifestation of the intuitive idea that slowing down the dynamics the system corresponds to effectively speeding up the time scale of the environment. Such frequency stretching, while irrelevant in the white noise limit, becomes relevant for time-correlated noise.

In the \ac{SchWARMA} formalism, there is not a mechanism for stretching pulses \textit{per se} as it operates at the gate level in a circuit (without pulse-level control on $H_c(t)$).  However, as discussed in the supplement to \cite{schultz2020schwarma}, it is possible to manipulate and stretch the spectrum of a \ac{SchWARMA} model. So, for the task of numerically simulating pulse stretching, instead of implementing equation Eq.\ \eqref{eq:pulse_stretching} one can simply implement Eq.\  \eqref{eq:spectrum_stretched} by directly transforming the spectrum of the \ac{SchWARMA} model.

\paragraph{Local unitary folding}
A possible way of effectively increasing the noise of a circuit is to insert after each noisy CNOT gate, the product of two additional CNOT gates \cite{dumitrescu2018cloud, He_2020_PRA}. In this way the ideal unitary is not changed, but the real dynamics are more noisy.
More generally, Ref.~\cite{giurgica2020digital} introduced  several digital noise scaling methods that are  based on the {\it unitary folding} replacement rule
\begin{equation}\label{eq:unitary-folding}
G \rightarrow G (G^\dag G)^n, \quad n = 0, 1, 2, \dots,
\end{equation}
where $G$ is a unitary operation associated to an individual gate.
If noise is absent, the replacement rule leaves the operation  unchanged since $G^\dag G$ is equal to the identity. On the contrary, if some base noise is associated to $G$, the unitary folding operation approximately scales the noise by an odd integer factor $\lambda = 1 + 2 n$. 

More precisely, by applying the unitary folding replacement to all the gates of an input circuit
\begin{equation}\label{eq:u-circuit}
    U = G_d G_{d-1} \dots G_1
\end{equation}
which is composed of $d$ gates $G_j$, we obtain new circuit $U'$ of depth $d' = (1 + 2n)d$ given by
\begin{equation}\label{eq:local-folding}
    U' = G_d (G_d^\dag G_d)^n G_{d-1} (G_{d-1}^\dag G_{d-1})^n \dots G_1(G_{1}^\dag G_{1})^n.
\end{equation}
The depth of the new circuit $U'$ is scaled by $\lambda = d'/d= 1 + 2n$ and, similarly, any type of noise which depends on the total number of gates will be effectively scaled by the same constant $\lambda$.  In Ref.~\cite{giurgica2020digital}, partial folding methods were proposed to obtain arbitrary real values of $\lambda$, but for simplicity in this work we only consider odd-integer scale factors.
We refer to~\eqref{eq:local-folding} as {\it local unitary folding}.

\paragraph{Global unitary folding}

Instead of locally folding all the gates, we can apply Eq.~\eqref{eq:unitary-folding} to the entire circuit. In this way, the circuit $U$ defined in Eq.~\eqref{eq:u-circuit} is simply mapped to
\begin{equation}\label{eq:global-folding}
    U' = U (U^\dag U)^n.
\end{equation}
Also in this case the total number of gates of the new circuit $U'$ is multiplied by $\lambda = d'/d= 1 + 2n$ corresponding to an effective scaling of the noise.

\paragraph{Gate Trotterization} 
In this work we also introduce another local noise-scaling method, acting at the level of individual gates, that we call {\it gate Trotterization} since it can be considered as a discretization of the continuous pulse-stretching technique. According to the gate Trotterization technique, each gate of the circuit is replaced as follows:
\begin{equation}\label{eq:gate-trotterization}
G \rightarrow \left(G^{1 / \lambda}\right)^\lambda, \quad \lambda = 0, 1, 2, \dots.
\end{equation}
For example, a Pauli $X$ rotation gate $R_X(\theta)$ is replaced by $\lambda$ applications of $R_X(\theta/\lambda)$.
Eq.~\eqref{eq:gate-trotterization} is similar to the local version of the unitary folding rule    \eqref{eq:unitary-folding} and, indeed, both methods replace a single gate with the product of $\lambda$ gates.
Compared to Eq.~\eqref{eq:unitary-folding}, the Trotter-like decomposition used in Eq.~\eqref{eq:gate-trotterization} is more uniform since equal elementary gates are used. On the other hand, 
a possible drawback of the gate Trotterization method is that $G^{1 / \lambda}$ may be compiled by the hardware in different ways depending on $\lambda$ and, therefore, the circuit depth may not get scaled as expected.

\section{Results}
\label{sec:results}

In the previous section, we defined several noise-scaling methods that can be used in zero-noise extrapolation.
In this section, we study how these different methods affect the performance of ZNE in the presence of time-correlated noise.
For all the simulations presented in this section we used the following Python libraries: Mezze~\cite{mezze} for modeling SchWARMA noise, Mezze's TensorFlow Quantum~\cite{broughton2021tensorflow} interface for simulating quantum circuits and Mitiq~\cite{larose2020mitiq} for applying unitary folding and zero-noise extrapolation. Code for specifying the circuits and the dephasing noise spectra used is also available at \cite{mezze}. 

\begin{figure}[t!]
    \includegraphics[width=1.0\columnwidth, left]{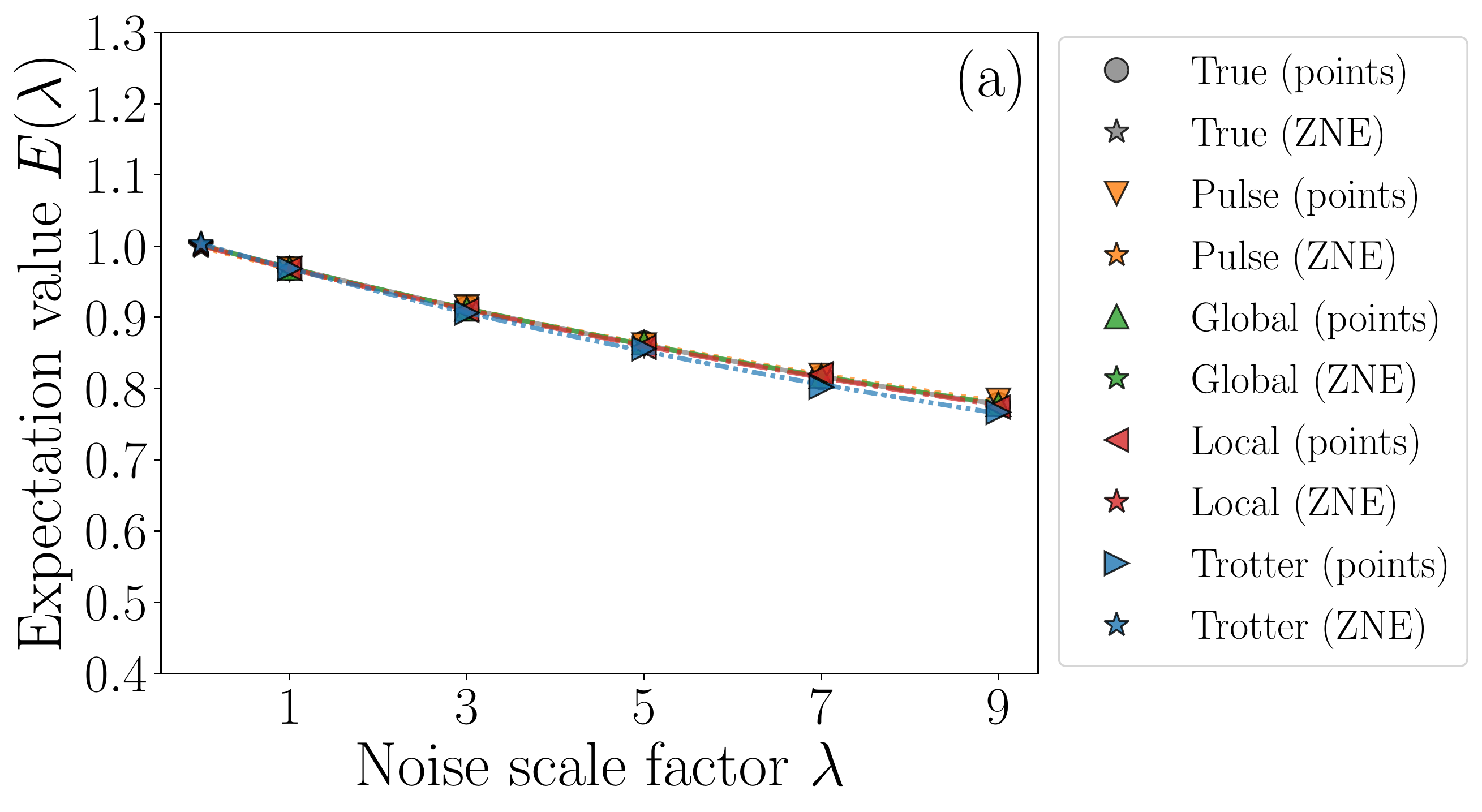}
    \includegraphics[width=0.81\columnwidth, left]{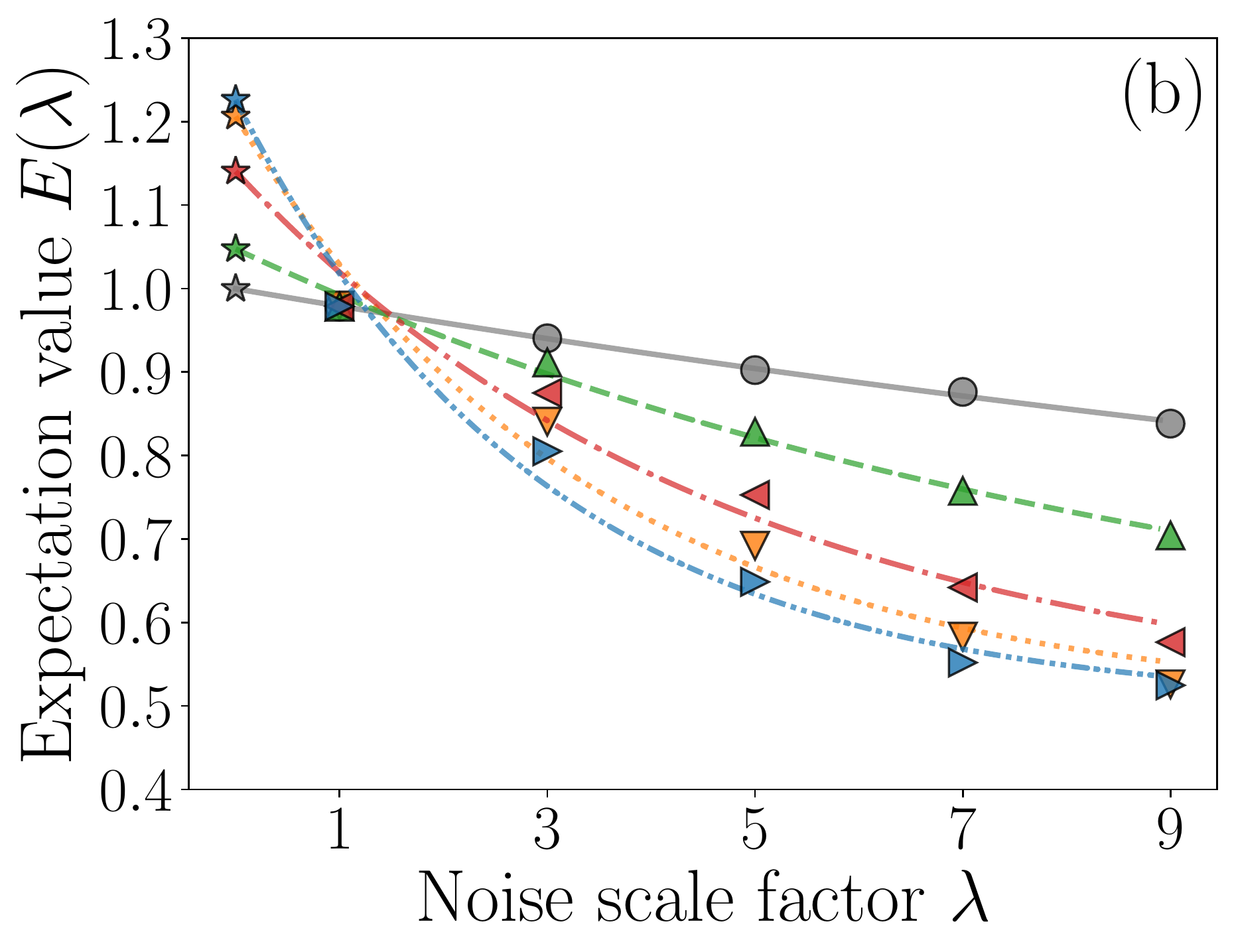}
    \caption{Comparison of different zero-noise extrapolations obtained with different noise scaling methods. We consider a representative (randomly generated) single-qubit randomized benchmarking circuit affected by dephasing noise of fixed integrated power. The two subfigures correspond to different noise spectra: (a) white noise, (b) $1/f$ pink noise. Both spectra are shown in Fig.~\ref{fig:noise_spectra}. 
    The expectation value $E(\lambda)=\Tr[ O \rho(\lambda)]$ is associated to the observable $ O=|0\rangle\langle0|$ measured with respect to the noise-scaled quantum state $\rho(\lambda)$.
    The colored triangles represent the noise-scaled expectation values; the dashed-dotted lines represent the associated exponential fitting curves; the colored {\it stars} represent the corresponding zero-noise extrapolations. The figure shows that
    the zero-noise limit obtained with global unitary folding (green star) is relatively close to the ideal
    result (gray star) even in the presence of strong time correlations in the noise. The ``true''  points (gray circles) are obtained by assuming the ideally scaled noise spectrum of Eq.~\eqref{eq:spectrum_scaled}, the associated fitting curve (solid gray line) produces the ideal zero-noise extrapolation (gray star).  }
    \label{fig:zne-comparison}
\end{figure}

\subsection{Zero-noise extrapolation with colored noise}

In this section we numerically simulate a simple ZNE experiment  with different noise scaling methods and with different noise spectra.   The results are reported in Fig.~\ref{fig:zne-comparison} and demonstrate the detrimental effect of time-correlated noise on ZNE. In Fig.~\ref{fig:zne-comparison}(a) the noise spectrum is white and all noise scaling methods  produce nearly identical expectation values. Correspondingly, the zero-noise limits (marked with stars in the plot) are nearly identical.
On the other hand, in Fig.~\ref{fig:zne-comparison}(b), the noise is colored (a $1/f$ ``pink'' spectrum) and different noise-scaling methods produce different expectation values. Correspondingly, the zero-noise limits (marked with stars in the plot) are also different.
This is the main qualitative result that this work aims to highlight:  compared to white noise, time-correlated noise can be much harder to mitigate via zero-noise extrapolation.

In the rest of this section, we study this aspect in a more quantitative way. In particular we study the performances of different noise-scaling methods for different types of noise spectra and different types of circuits.

\subsection{Comparing noise scaling methods}

\begin{figure}
    \centering
    \includegraphics[width=\columnwidth]{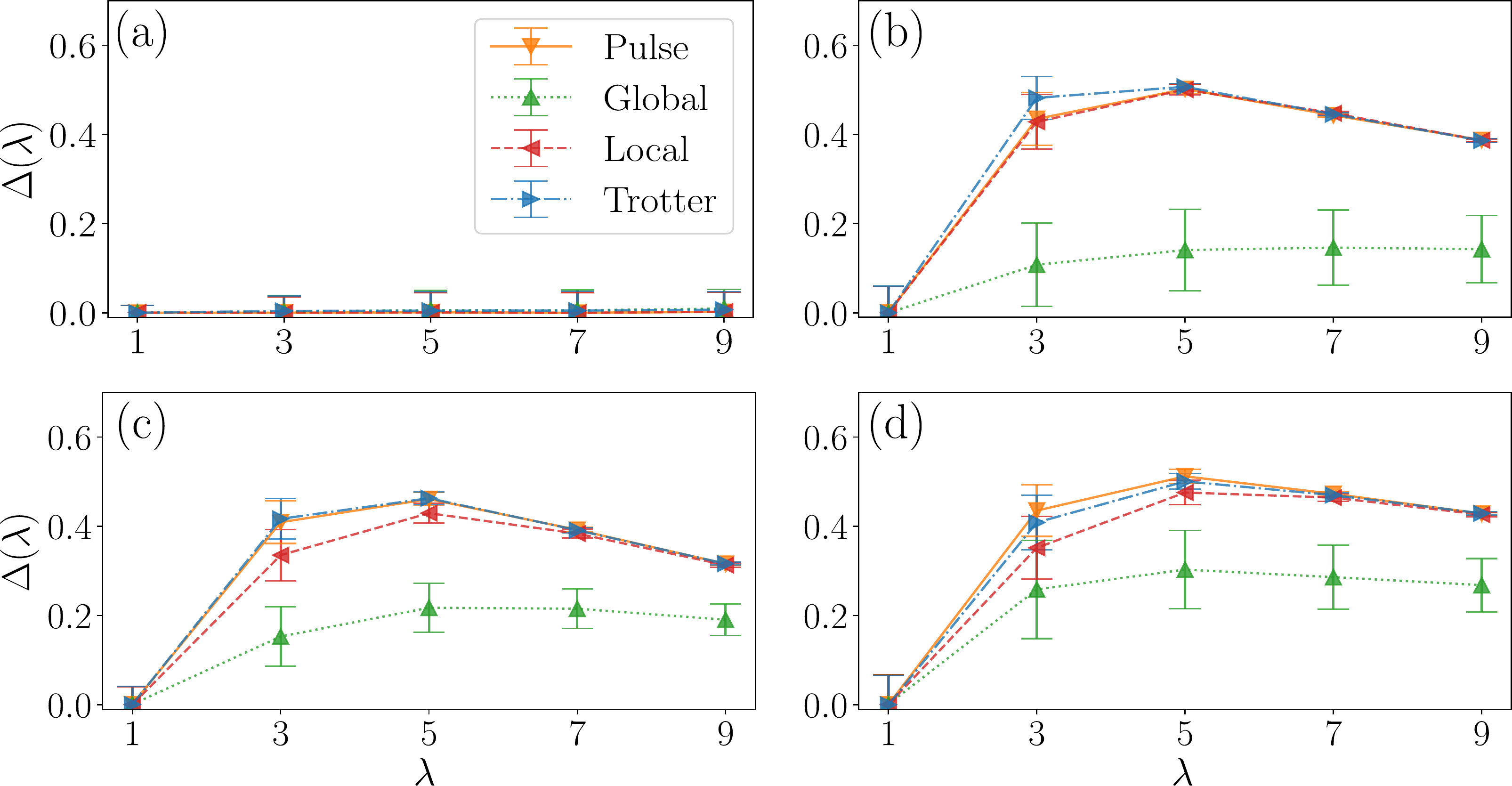}
    \caption{Average relative errors in noise scaling two-qubit randomized benchmarking circuits with (a) white noise, (b) lowpass noise, (c) $1 / f$ noise, and (d) $1 / f^2$ noise. Panel (a) shows no significant difference in scaling methods under white noise (no time correlations). Panels (b)-(d) show that global scaling is the lowest-error digital scaling method. The two-qubit randomized benchmarking circuits used here have, on average, 27 single-qubit gates and five two-qubit gates. For each circuit execution, $3000$ samples were taken to estimate the probability of the ground state as the observable. Points show the average results over fifty such circuits and error bars show one standard deviation.}
    \label{fig:2qrb_sim}
\end{figure}

Observing Fig.~\ref{fig:zne-comparison}(b) we notice that, at least for the particular circuit considered in the example, some noise scaling methods perform better than others in the presence of time-correlated noise.
In particular the extrapolation based on the global folding technique produces a relatively good approximation of the ideal result even in the presence of time-correlated noise.

To better investigate this phenomenon, we consider the \emph{relative noise-scaling error}
\begin{equation} \label{eq:relative-error}
    \Delta(\lambda) := \left| \frac{E(\lambda) - E^*(\lambda)}{E^*(\lambda)} \right|,
\end{equation}
as a figure of merit. Here, $E(\lambda)$ is the expectation value of interest evaluated with some particular noise scaling method and scale factor $\lambda$, and $E^*(\lambda)$ is the expectation value simulated with a noise spectrum ideally scaled according to Eq.~\eqref{eq:spectrum_scaled}.
In Fig.~\ref{fig:2qrb_sim} we plot the relative  error defined in Eq.~\eqref{eq:relative-error} for each noise-scaling method, after averaging the results over multiple instances of two-qubit randomized-benchmarking circuits. Here the expectation value of the observable $O = |00\rangle\langle 00|$ is considered.
The results of Fig.~\ref{fig:2qrb_sim} are consistent with those of Fig.~\ref{fig:zne-comparison} discussed  in the previous subsection. In fact, even after averaging over multiple random circuits, we observe that in the presence of white noise all noise scaling methods are practically equivalent to each other and are characterized by a small relative noise-scaling error. However, for all colored noise spectra, global folding is optimal when compared to other noise scaling methods.

\begin{figure}
    \centering
    \includegraphics[width=\columnwidth]{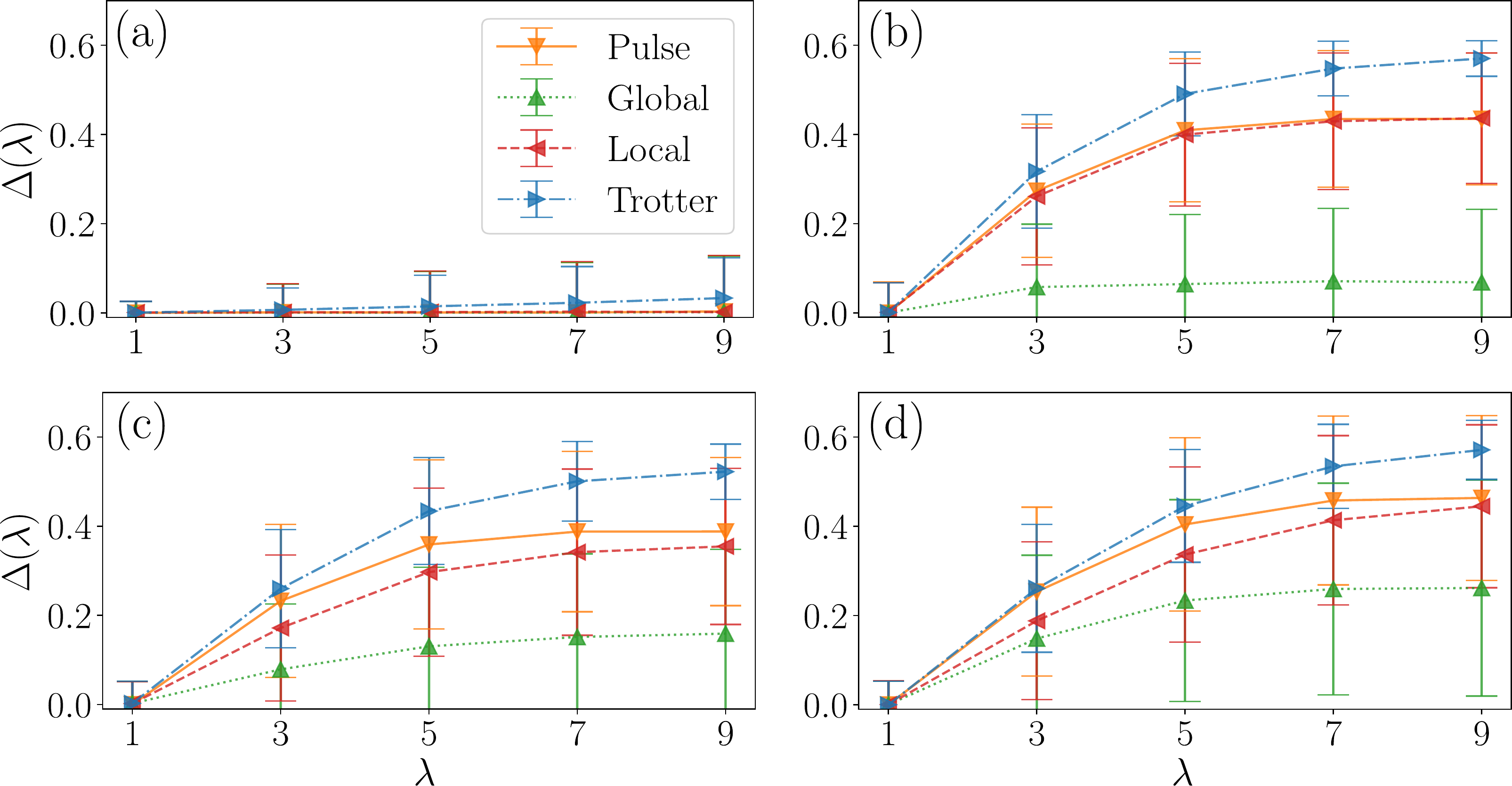}
    \caption{Relative errors in noise scaling two-qubit mirror circuits with (a) white noise, (b) lowpass noise, (c) $1 / f$ noise, and (d) $1 / f^2$ noise. Panel (a) shows no significant difference in scaling methods under white noise (no time correlations). Panels (b), (c) and (d) show global scaling is optimal with time-correlated noise. The two-qubit mirror benchmarking circuits used here have, on average, 26 single-qubit gates and eight two-qubit gates. For each circuit execution, $3000$ samples were taken to estimate the probability of the correct bitstring (defined by the particular mirror circuit instance) as the observable. Points show the average results over fifty such circuits and error bars show one standard deviation.}
    \label{fig:mirror}
\end{figure}

We repeat the same experiments using mirror circuits~\cite{Proctor_2020} and QAOA-like circuits instead of RB circuits. The former provides another type of randomized circuit structure used for benchmarking, and the latter provides a structured circuit.
Fig.~\ref{fig:mirror} shows the results using two-qubit mirror circuits. These circuits have 26 single-qubit gates and eight two-qubit gates on average. As with the randomized benchmarking circuits, 3000 samples were taken when executing each circuit to estimate the probability of sampling the correct bitstring. As shown in Fig.~\ref{fig:mirror}, the conclusion that global unitary folding most closely matches true noise scaling holds on average for mirror circuits as well. These results were averaged over fifty random mirror circuits.

\begin{figure}
    \centering
    \includegraphics[width=\columnwidth]{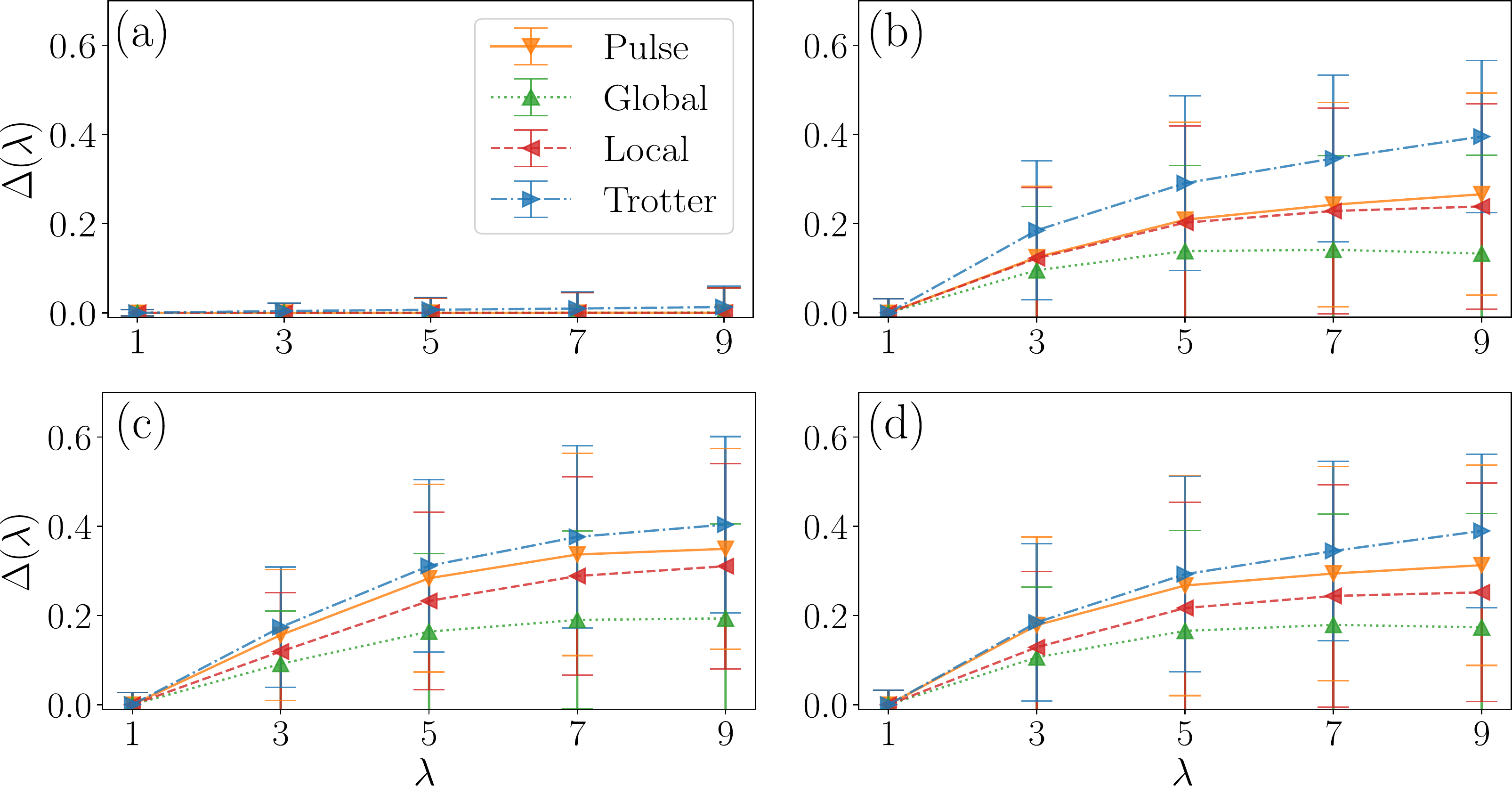}
    \caption{Relative errors in noise scaling two-qubit $p = 2$ QAOA circuits with (a) white noise, (b) lowpass noise, (c) $1 / f$ noise, and (d) $1 / f^2$ noise. Panel (a) shows no significant difference in scaling methods under white noise (no time correlations). Panels (b), (c) and (d) show global scaling is optimal with time-correlated noise. The two-qubit $p = 2$ QAOA circuits used here have eight single-qubit gates and four two-qubit gates. For each circuit execution, $3000$ samples were taken to estimate the probability of the ground state as the observable. (Note that the QAOA circuit $U$ is echoed such that the total circuit is $U U^\dagger = I$ without noise.) Points show the average results over fifty such circuits and error bars show one standard deviation.}
    \label{fig:qaoa}
\end{figure}

Fig.~\ref{fig:qaoa} shows the same experiment using QAOA circuits. These $n = 2$ qubit circuits have $p = 2$ QAOA rounds using the standard mixer Hamiltonian $H_M = \sum_{i = 1}^{n} \sigma^x_i$ and driver Hamiltonian $H_C = \sum_{ij} \sigma^z_i \sigma^z_j$. Denoting this circuit as $U$, we append $U^\dagger$ such that the final noiseless state is $|00\rangle$ independent of the randomly chosen angles $\mathbf{\beta}$ and $\mathbf{\gamma}$. A total of fifty circuits with random angles were simulated for the final results, again using 3000 samples to estimate the ground state probability for each circuit execution. The results in Fig.~\ref{fig:qaoa} have the highest variance of the three circuit types, but on average we still see that global unitary folding is closest to true noise scaling out of all scaling methods considered.

The conclusions of this subsection suggest that, even for different types of circuits, the effect of time-correlated noise on noise scaling methods is qualitatively similar. This intuition is consistent with the theoretical discussion presented in the next section, in which the performances of noise scaling methods are linked to their effective frequency modulation effects.

We emphasize that the comparison considered in this work is focused on one particular figure of merit: the robustness of a noise scaling method with respect to time-correlated  noise. Our results suggest that global folding outperforms the other methods considered with respect to this specific figure of merit. In a real-world scenario, the optimal noise-scaling method should be determined according to a more general cost-benefit analysis, e.g., taking into account the sampling cost, coherence time, and other hardware limitations. For instance, it may not be possible to use global noise scaling if the circuit length is comparable to the coherence time of the computer; in such circumstances, pulse stretching can amplify errors via small scale factors~\cite{kim_scalable_2021}, although potentially inaccurately in the presence of time-correlated noise as we have shown in this section.

\section{Discussion and Physical Interpretation}

In classical signal processing and control theory, the frequency response of a linear circuit or filter is used to understand how a circuit interacts with its input in the frequency domain. Using frequency domain techniques, one can understand and design filters that amplify (or pass through) frequencies that have signal content while attenuating frequencies that contain only noise.  In what follows, we introduce an analogous concept for quantum circuits that allows us to approximate the fidelity of a circuit subjected to dephasing noise. This circuit frequency response indicates where a given circuit is particularly sensitive or insensitive to noise in a given frequency range. We then analyze the impact of the various noise scaling techniques on this circuit frequency response to interpret the results of the previous section.

\subsection{Frequency response of a circuit}
%
The natural extension of the frequency response of a circuit to the quantum context is the so-called filter function formalism \cite{cywinski2008:fff,PhysRevLett.113.250501}. Details of the specific approach used here for multi-qubit, spatiotemporally correlated dephasing noise can be found in Appendix~\ref{app:ffs}, but the gist of the technique is that a circuit on $n$ qubits of time duration $T$ defines a set of real-valued switching functions 
\begin{equation}
    f_{\alpha\beta}(t)=\frac{1}{N}\Tr\left[U_0(T,t)A_\alpha U_0^\top(T,t)A_\beta\right]\,,
\end{equation}
defined by the action of a circuit's reverse-time propagator $U_0$ (see Eq.~\eqref{eq:U0}) on a set of traceless, Hermitian operators $\{A_\alpha\}$ that satisfy $\langle A_\alpha,A_\beta\rangle = \frac{1}{N}\Tr[A_\alpha A_\beta]=\delta_{\alpha,\beta}$ where $N=2^n$ (typically, the $A_\alpha$ are multi-qubit Pauli matrices). 

The Fourier transforms $F_{\alpha\beta}(\omega,T)=\int_0^Tdt\,f_{\alpha\beta}(t)e^{i\omega t}$ of these switching functions are used to define filter functions 
\begin{equation}
    \mathcal{F}_{\alpha\beta,\alpha^\prime\beta^\prime}(\omega,T) = \Re \left[F_{\alpha\beta}(\omega,T)F_{\alpha^\prime\beta^\prime}(-\omega,T)\right].
\end{equation}
Similar to classical frequency domain analysis, these filter functions interact with the dephasing noise spectra $S_{\alpha,\alpha'}(\omega)$ in a multiplicative fashion. Their product forms the integrand of the so-called overlap integral which is a key component of the second cumulant $C_O^{(2)}$:
\begin{eqnarray}\label{eq:cum2}
    \frac{\mathcal{C}^{(2)}_O(T)}{2} = \sum_{\alpha,\beta,\alpha^\prime,\beta^\prime}\int^{\infty}_{0} \frac{d\omega}{2\pi} S_{\alpha,\alpha^\prime}(\omega)\mathcal{F}_{\alpha\beta,\alpha^\prime\beta^\prime}(\omega,T)\mathcal{A}_{\beta\beta^\prime}.\nonumber\\
\end{eqnarray}
The overlaps between the noise power spectrum and filter functions scale operators $\mathcal{A}_{\beta\beta'}$ that are dependent on the observable $O$, see Eq.~\eqref{eq:Aop}.  The magnitude of $C_O^{(2)}$ (and thus the overlaps) can then be used to approximate the expectation of the noisy observable $\overline{\langle O\rangle}\approx \Tr[\exp(-C_0^{(2)}/2)\rho_{S,0}(T)O]$, where $\rho_{S,0}(T)$ is the final state of the ideal noiseless circuit, see discussion around Eq.~\eqref{eq:expO} in Appendix~\ref{app:ffs}.

The expression in Eq.~\eqref{eq:cum2} captures potential cross correlations in noise, but here since we consider independent $\sigma^z$-dephasing noise on each qubit, $S_{\alpha,\alpha'}=0$ when $\alpha\neq\alpha'$ and $\alpha$ is not a $\sigma^z$ operator on a given qubit. Furthermore, for the examples below we  compute the filter functions using instantaneous gates as specified by a circuit, but these expressions hold for piecewise constant controls to accommodate pulse shaping. In the context of noise scaling experiments, Eq.~\eqref{eq:cum2} provides a mechanism for understanding how the different noise scaling techniques perturb the filter functions to impact the resulting scaled expectations and thus the extrapolation process.

\begin{figure}
    \centering
    \includegraphics[width=.9\columnwidth]{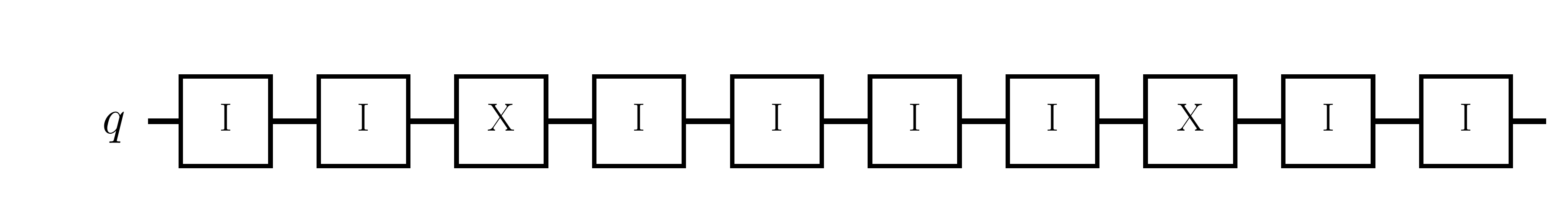}\\
    \includegraphics[width=.9\columnwidth]{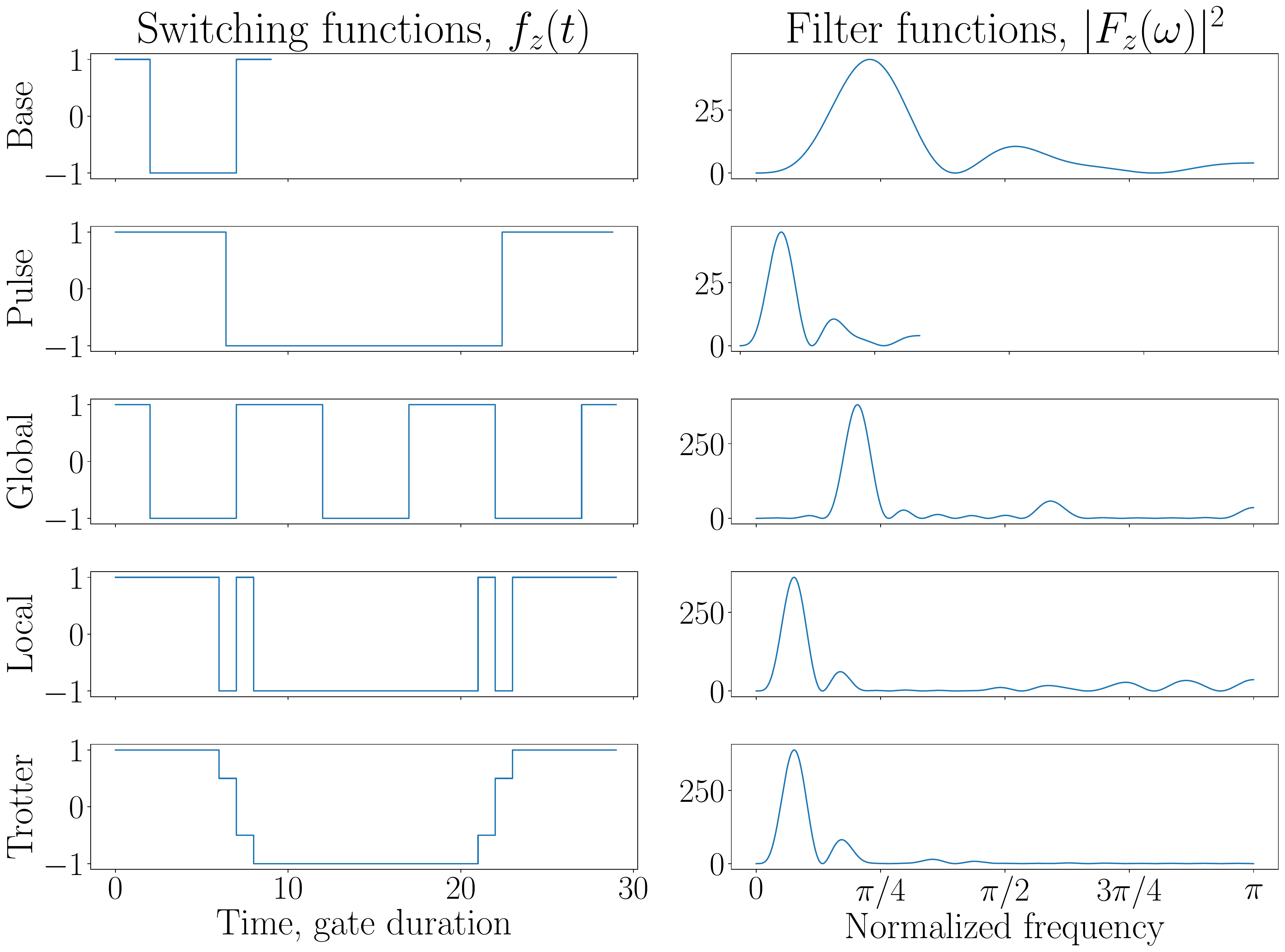}
    \caption{Filter function analysis of a \ac{CPMG} circuit. Top: \ac{CPMG} circuit with a delay of 2 gate times. Bottom: Plots of switching functions and filter functions of the \ac{CPMG} circuit and its various scaled versions.}
    \label{fig:simple_mod}
\end{figure}

The calculations for the multi-qubit case are quite involved, so in order to build better intuition we will also consider the simpler case of a single qubit subject to dephasing noise and single-axis $\sigma^x$ control, i.e., $H(t)=\eta(t)\sigma^z+\Omega(t)\sigma^x$. For this case, the filter function formalism can be recast in terms of a single \textit{complex} switching function $f_z(t)=\exp\left(-i\int_{0}^td\tau\,\Omega(\tau)\right)$ with Fourier transform $F_z(\omega)$. The filter function $\mathcal{F}_{zz}(\omega)=|F_z(\omega)|^2$ and power spectrum $S_{z,z}(\omega)=S_\eta(\omega)$ define the overlap integral
%
%
\begin{equation}
    \chi=\int_{-\infty}^{\infty}\frac{d\omega}{2\pi} S_{\eta}(\omega)|F_z(\omega)|^2\,.
\end{equation}
The overlap integral can be used to derive an approximation to the expectation of the noisy circuit states $\rho$ and an observable $O$, via 
\begin{equation}\label{eq:ffprob}
    E[\Tr[\rho\, O]]\approx A+B\exp(-\chi)\,,
\end{equation}
where $A$ and $B$ are functions of the ideal final state and observable $O$. 
%
An example \ac{CPMG} \cite{carr1954effects,meiboom1958modified} circuit and its corresponding $f_z$ and $|F_z(\omega)|^2$ are shown in Fig.~\ref{fig:simple_mod}. 

%


\subsection{Spectral analysis of noise scaling methods}
%

First, we will consider the simpler case of a single qubit subject to dephasing noise and single-axis $\sigma^x$ control, with the \ac{CPMG} circuit in Fig.~\ref{fig:simple_mod} serving as our canonical example. Using the filter function prediction from Eq.~\eqref{eq:ffprob} we have that direct noise scaling produces states $\rho_{dir}(\lambda)$ with expectation
\begin{equation}\label{eq:Ep_dir}
    E[\Tr[\rho_{dir}(\lambda) \, O]]\approx A+B\exp(-\lambda \chi)\,,
\end{equation}
where $\chi$ is the overlap integral of the base circuit. Note that direct noise scaling does not affect the circuit itself, and thus its switching and filter functions are unchanged. Similarly, following Eq.~\eqref{eq:pulse_stretching}, we have that pulse stretching produces the expectation 
\begin{multline}\label{Eq:Ep_pul}
    E[\Tr[\rho_{pul}(\lambda) \, O]]\approx\\ A+B\exp\left(-\lambda\int_{-\infty}^{\infty}\frac{d\omega}{2\pi} S_{z,z}(\omega/\lambda) |F_z(\omega)|^2\right)\,,
\end{multline}
with similar expressions for Eq.~\eqref{eq:cum2}, which is clearly not equal to Eq.~\eqref{eq:Ep_dir} in general. Equivalently, stretching the pulse amounts to stretching the switching functions and thus ``compressing'' a filter function response by a factor of $\lambda$. This shifts the filter function to lower frequencies, and thus the overlap with low-frequency noise will likely increase by a factor greater than $\lambda$. An example of the impact of pulse stretching on a \ac{CPMG} circuit is shown in Fig.~\ref{fig:simple_mod}, showing that the switching function is perfectly scaled in time, resulting in the corresponding frequency compression.

Gate Trotterization is similar in spirit to pulse stretching, but performed ``digitally.'' However, repeating a gate's control waveform $\lambda$ times with amplitude $1/\lambda$ is in general different from stretching a gate's control waveform (except in the case of rectangular pulses). Fig.~\ref{fig:simple_mod} shows a similar qualitative impact of gate Trotterization on the filter function as pulse stretching, in that the filter function is compressed to the low frequencies.  Since, unlike pulse stretching, the switching function now has intermediate values between $\pm1$ the filter function is distorted and not a ``perfect'' compression.

Like pulse stretching and gate Trotterization, local folding also increases the proportion of the filter function that overlaps with low frequency noise. However unlike pulse stretching and gate Trotterization, local folding also appears to generate response at high frequency.  Qualitatively, local folding ``pulls'' the filter function to the extreme frequencies from the middle of the spectrum. This behavior can be interpreted from the switching function, which now has (brief) oscillations whenever the original switching function had a change, and otherwise remains constant, see Fig.~\ref{fig:simple_mod}. These oscillations increase the high-frequency content of the filter function, whereas the longer duration of constant values increase the low-frequency content. Explicit filter function calculations for local folding can be found in Appendix~\ref{app:local} that generalize these observations to the multi-qubit case. With these general trends, we would again expect that the overlap integrals produced would not be particularly close to direct noise scaling.

Of the noise scaling methods studied, it appears that global folding preserves the most structure from the unscaled filter function.  The  circuit response shown in Fig.~\ref{fig:simple_mod} shows that scaling preserves the qualitative shape of the base circuit's filter function, but in accordance with well known results about \ac{CPMG} sequences, the frequency response is sharpened as it is repeated.  Qualitatively, it looks like the impact of global folding serves to ``resolve'' a coarse frequency response of the base circuit. Explicit calculations of the filter function (see Appendix~\ref{app:global}) show that the scaled portion of the circuit dominates the filter function response and approach a common, nontrivial limit. Thus, scaling in this case preserves some structure and produces overlap integrals that are somewhat close to direct noise scaling.

\begin{figure}
    \includegraphics[width=\columnwidth]{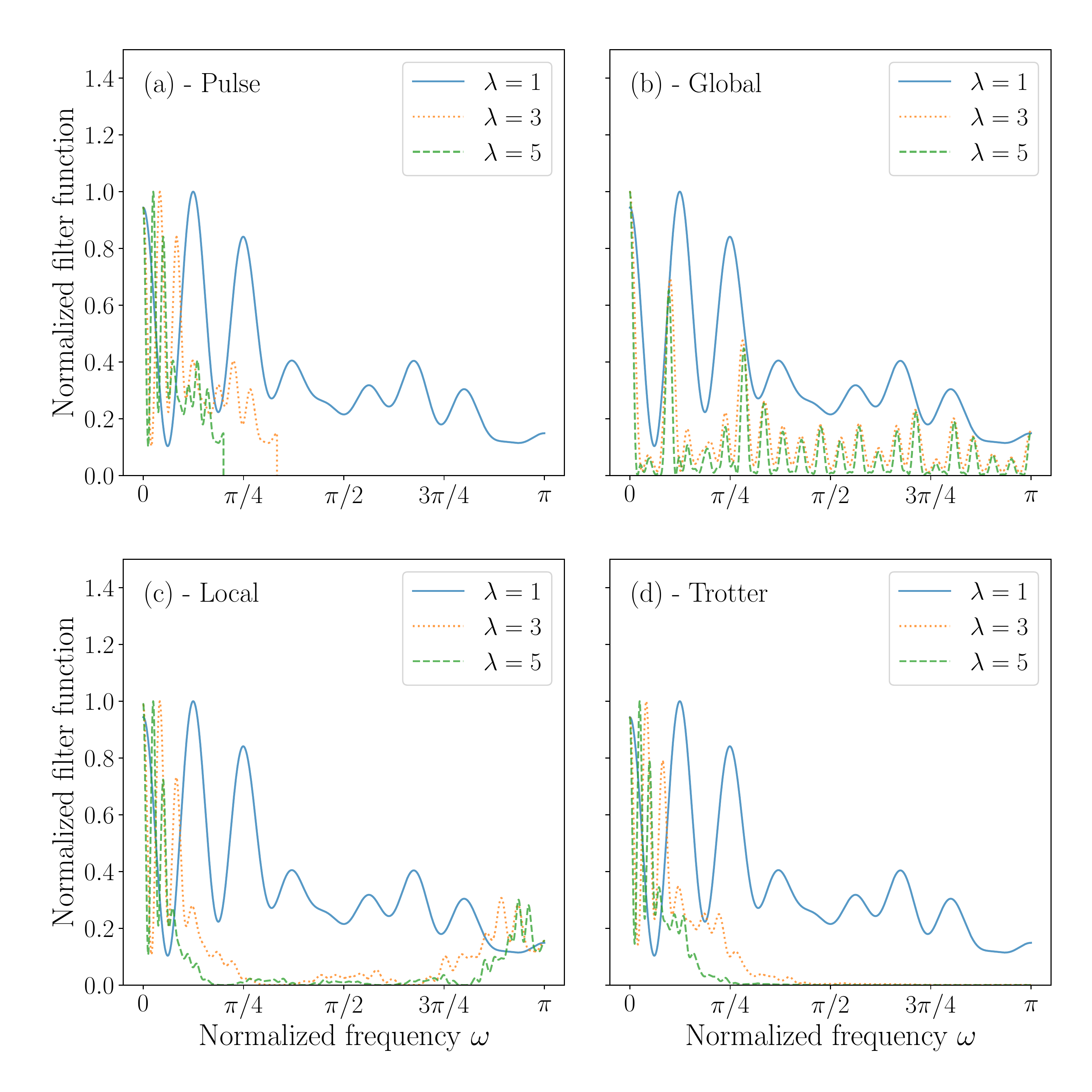}
    \caption{Largest magnitude filter function  of a two-qubit randomized benchmarking circuit of 
    Clifford depth 2 (actual depth 17) for different scale factors $\lambda$. All filter functions are normalized by their maximum values (otherwise the integral of the filter function scales by $\lambda$). Different subplots correspond to different noise scaling methods. All noise scaling methods change the frequency response of the circuit, however, global folding tends to preserve the qualitative shape of response function and, for this reason,
    it gives better performances for zero-noise extrapolation with colored noise.}
    \label{fig:frequency_response}
\end{figure}

The \ac{CPMG} sequence considered in the above discussion was chosen as an intuitive example for its well-studied frequency response \cite{alvarez2011:qns}, as well as clarity of exposition. However, the \ac{ZNE} simulations considered here use multi-qubit random circuits whose frequency response is less well studied. For these more complex circuits, we continue to see the same general trends in the filter function responses, as shown in Fig.~\ref{fig:frequency_response}. These circuits are longer and have greater gate density than the \ac{CPMG} example, and as such produce switching functions with many transitions that in turn leads to filter functions with many peaks and valleys. The spectral trends for the pulse stretching, local folding, and gate Trotterization methods in Fig.~\ref{fig:frequency_response} are quite clearly consistent with the \ac{CPMG} example, and in particular all exhibit increasing low frequency concentration as $\lambda$ increases (in addition to high frequency concentration for local folding). On the other hand, global folding appears to be approaching some limit that at least somewhat resembles the initial distribution of the frequency response (and can be assessed analytically -- see Appendix~\ref{app:global}). We interpret this as generalization of the sharpening of the spectral features well known for CPMG sequence, and multiple peaks are resolved from initially broad peaks as $\lambda$ increases.

These observations in the different noise scaling strategies explain the trends in Figs.~\ref{fig:zne-comparison} and \ref{fig:2qrb_sim}. As global folding produces scaled filter functions that best preserve the general balance across different frequency ranges, the overlap integrals of the globally folded circuits are the closest to the ideal scaling produced by direct noise scaling. The remaining three scaling approaches all produce some level of concentration at low frequencies, and thus tend to have much greater overlap with the low-frequency noise here.  As the pulse stretching and gate Trotterization approaches are very similar in spirit, they produce similar extrapolations. Furthermore, unlike local folding, these two approaches have all their concentration at low frequency, thus producing the most overlap leading to the worst extrapolation error. Local folding, which includes some high frequency content (based on the proportion of the original circuit's frequency response above $\pi/2$), produces overlaps that lie between the global folding and the stretching/Trotterization approaches.

We note that the trends observed above and the intuition behind them is a direct consequence of the correlated noise classes considered, all of which are fundamentally low frequency.  Thus, pulse stretching, gate Trotterization, and local folding produce larger overlaps with the low-frequency noise and drastically bias the noise extrapolation process.  In contrast, if the noise was band limited (say between $\pi/4$ and $3\pi/4$ in normalized frequency) we would expect that global folding would continue to track direct noise scaling the best.  However, analysis of the other three techniques would be challenging as the overlap integral with these would essentially vanish as the scaling increased. Without knowing the true expectation and the underlying noise spectra, it would be unclear if the leveling out of the scaled expectation values would be due to the overlap integral approaching infinity (i.e., too much noise) or vanishing (i.e., decoupling from the noise).  Similarly, if the noise were purely high frequency, we would expect the pulse stretching and gate Trotterization approaches to be insensitive, local folding method to be more sensitive, and global folding between them. Finally, extremely narrow band noise could potentially lie in a ``valley'' in the scaled response (obviously this is circuit dependent), and thus overlap integrals would vanish for all the noise scaling approaches considered here.

\section{Conclusion}

In this work, we have demonstrated the effect of time-correlated noise on zero-noise extrapolation. Using the \ac{SchWARMA} technique to model time-correlated dephasing noise, we presented the results of several numerical experiments showing that global unitary folding produces the lowest error relative to direct noise scaling. We analyzed our observed results and provided a physical interpretation in terms of the spectral analysis of the considered noise scaling methods.

Noise injection as a method for noise scaling is a theoretically ideal mechanism for noise scaling, but given the limitations on estimating, emulating, and injecting the native spectra and noise mechanism(s), this is not likely feasible in most situations. On the other hand, global noise folding is broadly applicable and our work suggests its use in global noise scaling for zero-noise extrapolation, if possible, whenever noise may be time-correlated. An obviously important consideration is which quantum computer architectures may have time-correlated noise, a question we do not explicitly consider in this paper, but do note that time-correlated noise has been widely observed in both research-grade qubit experiments and cloud-based NISQ systems, in a variety of platforms \cite{bylander2011:fnoise, yan2013:fnoise,burnett2014:fnoise,muller2015:fnoise,burnett2019:fnoise,basset2014:fnoise,chan2018:qns,struck2020:fnoise, pokharel2018demonstration,zhang2021predicting,murphy2021universal,tripathi2021suppression,niu2022pulse}. We note that global folding is not the only possible noise scaling method suitable for time-correlated noise: other methods could be defined and analyzed, e.g., folding the first half and second half of the gates in a unitary separately. Our work provides the theoretical and practical tools to analyze the performance of such methods under a wide variety of noise models.

\section*{Data availability}

Software for reproducing all numerical results is available at \href{https://github.com/mezze-team/mezze}{https://github.com/mezze-team/mezze}.

\section*{Acknowledgements}
We thank Sarah Kaiser for helpful discussions.
This work was supported by the U.S. Department of Energy, Office of Science, Office of Advanced Scientific Computing Research, Accelerated Research in Quantum Computing under Award Number DE-SC0020266 and DE-SC0020316. RL acknowledges support from a NASA Space Technology Graduate Research Fellowship.

\bibliography{refs}

\appendix

\section{Consistency between different theories of pulse-stretching} \label{app:consistency}

Our work is based on a semiclassical theory of time-correlated noise, according to 
which, the pulse-stretching technique induces two effective changes   on the noise spectrum: (i) it scales the noise level by a constant $\lambda$, (ii) it also stretches the noise spectrum on the frequency axis by the same constant. Both effects are formally summarized in Eq.~\eqref{eq:spectrum_stretched} derived in the main text.

In Ref.~\cite{temme2017error}, a different formalism, based on a master equation with a time-independent noise operator, was used to study the pulse-stretching technique.
More precisely, a system evolving according to the following master equation was considered:
\begin{equation} \label{eq:master-equation-temme}
    \frac{\partial}{\partial t} \rho(t) = - [K(t), \rho(t)] + \mathcal L (\rho(t)),
\end{equation}
where $K(t)$ is the system Hamiltonian and $\mathcal L$ is a time-independent noise super-operator. As shown in Ref.~\cite{temme2017error}, the effect of pulse stretching (i.e., $K(t) \longrightarrow 1/\lambda K(t/\lambda)$) is equivalent to an effective master equation:
\begin{equation} \label{eq:stretched-master-equation-temme}
    \frac{\partial}{\partial t'} \rho(t') = - [K(t'), \rho(t')] +  \lambda \mathcal L (\rho(t')),
\end{equation}
where $t'=\lambda t$. In practice pulse-stretching induces a multiplicative scaling of the noise operator $\mathcal L \longrightarrow \lambda \mathcal L$.

The master equation Eq.~\eqref{eq:master-equation-temme} is typically used to model Markovian noise (no time-correlations). In this case, the Hilbert space of the environment can be traced out such that $\rho$ represents the reduced state of the system evolving  according to the master equation Eq.~\eqref{eq:master-equation-temme}.
In this white-noise regime, also our semiclassical theory of pulse-stretching predicts a simple multiplicative scaling of noise power and this is indeed consistent with Eq.~\eqref{eq:stretched-master-equation-temme}.

What happens for a non-Markovian environment with a colored noise spectrum? In this case,  our semiclassical theory suggests that pulse-stretching induces, in addition to a multiplicative scaling, also a scaling of the frequency axis of the noise spectrum (see Eq.~\eqref{eq:spectrum_stretched}). This may seem to contradict 
the simple multiplicative scaling of the noise $\mathcal L \longrightarrow \lambda \mathcal L$ derived in Ref.~\cite{temme2017error} and reported in Eq.~\eqref{eq:stretched-master-equation-temme}. However, as explained below, both theoretical derivations are actually consistent with each other.

In principle, the master equation \eqref{eq:master-equation-temme} can be used to model a non-Markovian bath by representing with $\rho$ the global quantum state (system + bath) instead of the reduced state of the system. In this global picture, a non-Markovian bath can be modeled by a time-independent noise operator  $\mathcal L(\rho)$ that includes an interaction Hamiltonian term $H_{\rm SB}$ and the bare Hamiltonian $H_{\rm B}$ acting on the bath only (see Supplemental Material of Ref.~\cite{temme2017error})

\begin{equation} \label{eq:non-markov-L}
     \mathcal L (\rho(t)) = - i [H_{\rm SB} +  H_{\rm B}, \rho(t) ],
\end{equation}
which we can split as the sum of two terms $\mathcal L = \mathcal L_{\rm SB} + \mathcal L_{\rm B}$, where $\mathcal L_{\rm SB}(\rho) = -i [H_{\rm SB}, \rho]$ and $\mathcal L_{\rm B}(\rho) = -i [H_{\rm B}, \rho]$.
In this case, the simple multiplicative scaling $\mathcal L \longrightarrow \lambda \mathcal L$ induced by the pulse-stretching technique according to Eq.~\eqref{eq:stretched-master-equation-temme} has actually two physically different effects: (i) $\mathcal L_{SB} \longrightarrow \lambda \mathcal L_{SB}$ corresponding to  a scaling of the noise power and (ii) $\mathcal L_{B} \longrightarrow \lambda \mathcal L_{B}$ corresponding to an effective scaling of the all the characteristic frequencies of the bath and, therefore, to a frequency stretching of the noise spectrum. These two effects are consistent with the semiclassical theory of pulse-stretching presented in this work and, in particular, with Eq.~\eqref{eq:spectrum_stretched}.

\section{Filter function formalism for quantum circuits}\label{app:ffs} 
Consider an $n$-qubit system governed by a Hamiltonian
\begin{equation}
    H(t) = H_0(t) + H_E(t),
\end{equation}
where $H_0(t)$ defines a sequence of control operations applied to the quantum system and $H_E(t)$ defines the error Hamiltonian. We will assume piecewise-constant evolution such that $H_0(t)=\sum_i s_i(t) H_i$, where $s_i(t)=1$ when $t\in[t_{i-1},t_i)$ and $s_i(t)=0$ otherwise. The resulting pure control evolution is
\begin{eqnarray}\label{eq:U0}
U_0(T,0) &=& \mathcal{T}_+ e^{-i\int^{T}_0 H_0(s) ds}\nonumber\\
&=& U_K U_{K-1} \cdots U_1 \nonumber\\
&=& U_{K:1},
\end{eqnarray}
with $U_j=e^{-i(t_j-t_{j-1})H_j}$ and $T$ designating the total circuit runtime. The last equality is defined in anticipation of the subsequent switching function calculations. 

The error Hamiltonian can include anything from systematic control noise to interactions between the system and its environment. Here, we focus on semiclassical, spatiotemporally correlated noise: $H_E(t)=\sum_\alpha b_\alpha(t)A_\alpha $. The operators $A_\alpha$ are Hermitian, traceless and form an operator basis on the system Hilbert space with respect to the Hilbert-Schmidt norm, i.e., $\braket{A_\alpha, A_\beta}=1/N\Tr[A_\alpha A_\beta]=\delta_{\alpha,\beta}$, with $N=2^n$ denoting the Hilbert space dimension. The noise couples to the system via $b_\alpha(t)$, which are defined as classical wide-sense stationary, Gaussian variables. Hence, the statistical properties of $b_\alpha(t)$ are characterized by the mean $\overline{b_\alpha(t)}$ and the two-point correlation functions $C_{\alpha\beta}(t_1-t_2)=\overline{b_\alpha(t_1)b_\beta(t_2)}$. Note that this model can be used to represent both additive control noise and interactions between the quantum system and a classical environment.

Assuming the weak noise limit, i.e., $\|H_E(t)\|T\ll1$, we can examine the dynamics generated by $H(t)$ perturbatively by moving into an interaction picture with respect to $H_0(t)$. This is performed by representing the full dynamics operator $U(T)=\mathcal{T}_+ e^{-i\int^{T}_0 H(s) ds}$ as $U(T)=\tilde{U}_E(T)U_0(T)$, where $\tilde{U}_E(T)=e^{-i\int^{T}_0 \tilde{H}_E(s) ds}$ with
\begin{eqnarray}
\tilde{H}_E(t) &=& U_0(T,t)H_E(t)U^\dagger_0(T,t)\nonumber\\
&=& \sum_\alpha U_0(T,t) A_\alpha U^\dagger_0(T,t) B_\alpha(t)\nonumber\\
&=& \sum_{\alpha,\beta} f_{\alpha\beta}(t) A_{\beta} b_\alpha(t).
\label{eq:switch-fun}
\end{eqnarray}
Note that the rotated error Hamiltonian is expressed in terms of a ``reverse" interaction picture with respect to the pure control evolution. The functions $f_{\alpha\beta}(t)$ are known as the switching functions and are defined by
\begin{equation}
    f_{\alpha\beta}(t) = \frac{1}{N}\Tr\left[U_0(T,t) A_\alpha U^\dagger_0(T,t) A_\beta\right].
\end{equation}
As we will see, the switching functions are an integral part of the filter function formalism.

The dynamics of an observable with respect to $U(T)$ can be written as
\begin{eqnarray}\label{eq:expO}
\overline{\braket{O}}&=&\overline{\Tr[\rho_S(T)O]}\nonumber\\
&=& \overline{\Tr[\tilde{U}_E(T)U_0(T)\rho_S(0)U^\dagger_0(T)\tilde{U}^\dagger_E(T)O]}\nonumber\\
&=& \Tr[\Lambda(T)\rho_{S,0}(T) O],
\end{eqnarray}
where the last equality conveniently illustrates the utility of the particular rotated frame chosen above. The term $\Lambda(T)=\overline{O^{-1}\tilde{U}_E(T)O\tilde{U}_E(T)}$ constitutes the error operator, while $\rho_{S,0}(T)=U_0(T)\rho_S(0)U^\dagger_0(T)$ defines the time-evolved state resulting from the noiseless circuit implementation. The error operator can be expressed as a cumulant expansion $\Lambda(T)=\exp\left(\sum^\infty_{n=1}(-i)^n\mathcal{C}^{(n)}_O(T)/n!\right)$ that can be truncated to second order if the noise is sufficiently weak and the time is sufficiently short~\cite{pazsilva2017:qns}. 

We will focus on zero-mean noise, and thus, $\overline{b_\alpha(t)}=0$ and $\mathcal{C}^{(1)}_O(T)=0$. As a result, the first non-zero term is $\mathcal{C}^{(2)}_O(T)$, which can be written as
\begin{eqnarray}
    \frac{\mathcal{C}^{(2)}_O(T)}{2} = \sum_{\alpha,\beta,\alpha^\prime,\beta^\prime}\int^{\infty}_{0} \frac{d\omega}{2\pi} S_{\alpha,\alpha^\prime}(\omega)\mathcal{F}_{\alpha\beta,\alpha^\prime\beta^\prime}(\omega,T)\mathcal{A}_{\beta\beta^\prime},\nonumber\\
\end{eqnarray}
The noise power spectral density $S_{\alpha\alpha^\prime}(\omega)$ is defined via
\begin{equation}
    C_{\alpha\alpha^\prime}(t_1-t_2)=\frac{1}{2\pi}\int^{\infty}_{-\infty}S_{\alpha\alpha^\prime}(\omega)\, e^{i\omega(t_1-t_2)} d\omega,
\end{equation}
while the filter functions are
\begin{equation}
    \mathcal{F}_{\alpha\beta,\alpha^\prime\beta^\prime}(\omega,T) = \Re \left[F_{\alpha\beta}(\omega,T)F_{\alpha^\prime\beta^\prime}(-\omega,T)\right].
\end{equation}
with $F_{\alpha\beta}(\omega,T)=\int^{T}_0 dt\, f_{\alpha\beta}(t)e^{i\omega t}$. The operator $\mathcal{A}_{\beta\beta^\prime}$ is defined as
\begin{eqnarray}\label{eq:Aop}
    \mathcal{A}_{\beta\beta^\prime} &=& A_{\beta}A_{\beta^\prime} - O^{-1}A_{\beta}OA_{\beta^\prime}- A_{\beta}O^{-1}A_{\beta^\prime}O\nonumber\\
    && + O^{-1}A_{\beta}A_{\beta^\prime}O
\end{eqnarray}
We will now use this representation to derive analytical expressions for the filter functions resulting from various unitary folding techniques.

\section{Filter function perspective on global folding}\label{app:global}
Consider a global folding protocol defined as $U_{GF}[(2M+1)T,0]=U_0(T)\left[U^\dagger_0(T)U_0(T)\right]^M$, where $U_0(t)$ is the unitary representing the desired quantum algorithm and $M$ is the number of folding repetitions to be performed. The total time required to implement the algorithm is denoted by $T$. 

\subsection{Switching functions}
The switching functions resulting from the global folding protocol are given by
\begin{equation}
    f_{\alpha\beta}(t)= \left\{
    \begin{array}{lcr}
    f^{(1)}_{\alpha\beta}(t) &:& t\in[0,2MT)\\  f^{(2)}_{\alpha\beta}(t) &:& t\in[2MT,(2M+1)T)
    \end{array}
    \right..
\end{equation}
We partition them into two terms: the first, denoted with superscript (1), is defined during the global folding as
\begin{equation}
f^{(1)}_{\alpha\beta}(t) = \frac{1}{N}\Tr\left[U_0(T)\Gamma(2MT,t)A_\alpha \Gamma^\dagger(2MT,t)U^\dagger_0(T)A_\beta\right].
\end{equation}
The operator $\Gamma(2MT,t)$ captures the partial (reverse) unitary dynamics of $[U^\dagger_0(T)U_0(T)]^M$. By again expanding into the operator basis of $A_\alpha$, we define an additional switching function
\begin{equation}
    y_{\alpha\gamma}(t) = \frac{1}{N} \Tr\left[\Gamma(2MT,t)A_\alpha\Gamma^\dagger(2MT,t)A_\gamma\right],
\end{equation}
such that
\begin{equation}
    \Gamma(2MT,t)A_\alpha \Gamma^\dagger(2MT,t) = \sum_\gamma y_{\alpha\gamma}(t)A_\gamma.
\end{equation}
As a result, we can express $f^{(1)}_{\alpha\beta}(t)$ as
\begin{equation}
    f^{(1)}_{\alpha\beta}(t) = \frac{1}{N}\sum_\gamma y_{\alpha\gamma}(t)\Tr\left[U_0(T)A_\gamma U^\dagger_0(T)A_\beta\right].
\end{equation}
The advantage of this representation is that we have now defined the switching function $f^{(1)}_{\alpha\beta}(t)$ in terms of a mirror symmetric switching function $y_{\alpha\gamma}(t)$. Mirror symmetric switching functions satisfy the property
\begin{equation}
    y_{\alpha\gamma}(t) = y_{\alpha\gamma}(2T-t),
    \label{eq:mirror_sym}
\end{equation}
which we will find useful when examining the filter functions of global folding. Lastly, the second term of the switching function $f_{\alpha\beta}(t)$ is defined after the global folding and given by
\begin{eqnarray}
f^{(2)}_{\alpha\beta}(t) &=& \frac{1}{N}\sum_j g_j(t)\times\nonumber\\
&&\Tr\left[U_{K:j+1}e^{-i(t_j-t)H_j}A_\alpha e^{i(t_j-t)H_j}U^\dagger_{K:j+1}A_\beta\right],\nonumber\\
\end{eqnarray}
with $g_j(t)=\Theta(t-t_{j-1})-\Theta(t-t_j)$, and $\Theta(t)$ denoting the Heaviside function.

\subsection{Filter functions}
The filter functions are defined via products of Fourier transforms of switching functions. Since the switching functions can be partitioned into two terms, the filter functions can be partitioned into four terms:
\begin{equation}
F_{\alpha\beta,\alpha^\prime\beta^\prime}(\omega,T)=\sum^{2}_{i,j=1} G^{(i,j)}_{\alpha\beta,\alpha^\prime\beta^\prime}(\omega,T)
\label{eq:partitioned-FF}
\end{equation}
where
\begin{widetext}
\begin{eqnarray}
    G^{(1,1)}_{\alpha\beta,\alpha^\prime\beta^\prime}(\omega,T)&=& \frac{\sin^2(M\omega T)}{\sin^2(\omega T)} F^{(1)}_{\alpha\beta}(-\omega,T)F^{(1)}_{\alpha^\prime\beta^\prime}(\omega,T)\\
    G^{(1,2)}_{\alpha\beta,\alpha^\prime\beta^\prime}(\omega,T)&=& e^{i(3M+1)\omega T}\frac{\sin(M\omega T)}{\sin(\omega T)} F^{(1)}_{\alpha\beta}(-\omega,T)F^{(2)}_{\alpha^\prime\beta^\prime}(-\omega,T)\\
    G^{(2,1)}_{\alpha\beta,\alpha^\prime\beta^\prime}(\omega,T)&=& [G^{(1,2)}_{\alpha\beta,\alpha^\prime\beta^\prime}(\omega,T)]^*\\
    G^{(2,2)}_{\alpha\beta,\alpha^\prime\beta^\prime}(\omega,T)&=& F^{(2)}_{\alpha\beta}(\omega,T)F^{(2)}_{\alpha^\prime\beta^\prime}(-\omega,T).
\end{eqnarray}
\end{widetext}
The component filter functions are determined by
\begin{eqnarray}
    F^{(1)}_{\alpha\beta}(\omega,T)&=&\int^{2T}_{0}dt\,f^{(1)}_{\alpha\beta}(t)e^{i\omega t}, \\
    F^{(2)}_{\alpha\beta}(\omega,T)&=&\int^{T}_{0}dt\,f^{(2)}_{\alpha\beta}(t)e^{i\omega t}.
\end{eqnarray}
Note that $G^{(1,1)}_{\alpha\beta,\alpha^\prime\beta^\prime}(\omega,T)$ and $G^{(1,2)}_{\alpha\beta,\alpha^\prime\beta^\prime}(\omega,T)$ exhibit ``comb-like" behavior conveyed by the presence of the quotient of sinusoidal functions. These factors appear from the $M$ repetitions of global folding, and they are responsible for the more distinct features in the filter function as $M$ increases.

\section{Filter function perspective on local folding}\label{app:local}
Local folding is generically described by the total unitary
\begin{eqnarray}
    U_{LF}(T,0)&=&U_K\left(U^\dagger_K U_K\right)^M\cdots U_1\left(U^\dagger_1 U_1\right)^M\nonumber\\
    &=& U_K\, \Gamma_K(T_K)\cdots U_1\,\Gamma_1(T_1),
\end{eqnarray}
where each gate is subject to a folding interval. Each local folding is equivalent in repetition, occurring $M$ total times for each of the gates $U_j$. It is assumed that each gate takes an equivalent amount of time $\tau$, and therefore the timing of each gate is given by
\begin{equation}
    t^{(j)}_k = [(2M+1)(j-1) + k]\tau,
\end{equation}
for $j=1,\ldots,K$ and $k=1,\ldots, 2M+1$. We define the local folding unitary $\Gamma_j(T_j) = (U^\dagger_j U_j)^M$, where the total time $T_j=t^{(j)}_{2M}$, for convenience and in anticipation of the subsequent calculations. At intermediate times, the folding operator is given by
\begin{equation}
    \Gamma_j(T_j,t) = \left\{
    \begin{array}{lcr}
         (U^\dagger_j U_j)^{M-k} e^{i(t^{(k)}_k - t) H_j} & :& k\,\, \textrm{odd} \\
         (U^\dagger_j U_j)^{M-k} U^\dagger_j e^{-i(t^{(j)}_k - t) H_j} & :& k\,\, \textrm{even}
    \end{array}
    \right..
\end{equation}
Note that we have defined the local folding operator with respect to the total time as this will naturally appear from the ``reverse" propagator $U_0(T,t)$; see Eq.~(\ref{eq:switch-fun}). Furthermore, note that for brevity we will use the notation $\Gamma_j(t) = \Gamma_j(T_j,t)$.

\subsection{Switching functions}
While local folding utilizes a more complex folding procedure than global folding, the switching functions can still be partitioned into folding and post-folding terms. Specifically, the folding terms reside in the domain $t\in[t^{(j-1)}_{2M+1}, t^{(j)}_{2M})$ and the post-folding terms can be defined within $t\in[t^{(j)}_{2M}, t^{(j)}_{2M+1})$. This construction allows for the switching function to be expressed as $f_{\alpha\beta}(t)=f^{(1)}_{\alpha\beta}(t)+f^{(2)}_{\alpha\beta}(t)$, where 
\begin{equation}
    f^{(\mu)}_{\alpha\beta}(t)=\sum_j g^{(\mu)}_j(t)f^{(\mu,j)}_{\alpha\beta}(t).
\end{equation} 
The indices $\mu=1,2$ characterize the folding and post-folding terms, respectively. The functions $g^{(1)}_j(t)=\Theta[t-t^{(j-1)}_{2M+1}]-\Theta[t-t^{(j)}_{2M}]$ and $g^{(2)}_j(t)=\Theta[t-t^{(j)}_{2M}]-\Theta[t-t^{(j)}_{2M+1}]$ capture the piecewise features of the gate folding periods and their final implementation.

During a local folding period, the switching function is defined by 
\begin{equation}
    f^{(1,j)}_{\alpha\beta}(t)=\frac{1}{N}\sum_\gamma y^{(j)}_{\alpha\gamma}(t)\Tr\left[Q_{K:j}\Gamma^{\dagger}_j(T_j) A_\gamma \Gamma_j(T_j)Q^\dagger_{K:j}A_\beta\right],
\end{equation}
where $Q_{k:j}=U_k\Gamma_k(T_k)\cdots U_j\Gamma_j(T_j)$. As in the global folding case, we introduce an additional switching function
\begin{equation}
    y^{(j)}_{\alpha\gamma}(t) = \frac{1}{N}\Tr[\Gamma_{j}(t)A_\alpha \Gamma^\dagger_{j}(t) A_\beta]
\end{equation}
which satisfies the mirror symmetric condition described in Eq.~(\ref{eq:mirror_sym}) for the $j$th local folding interval. Once again, this property will prove useful during the calculation of the filter functions.

Local folding operations are followed by the implementation of the folded gate. The switching function describing this post-folding period is
\begin{eqnarray}
    f^{(2,j)}_{\alpha\beta}(t)=\frac{1}{N}\Tr\left[Q_{K:j+1} e^{-i(t^{(j)}_{2M+1}-t)H_j}A_\alpha \times \right. \nonumber\\
    \left. Q^\dagger_{K:j+1} e^{i(t^{(j)}_{2M+1}-t)H_j}A_\beta\right].
\end{eqnarray}
Partial time evolution during $t\in[t,t^{(j)}_{2M+1})$ captures the switching function dynamics during the gate. Note that this contribution disappears if the gates are assumed to be instantaneous.

\subsection{Filter functions}
There are four primary filter function that characterize the self-interference and cross-interference between local folding and post-folding intervals. While this is a similar construction to the global folding case in that the filter function $F^{(i,j)}_{\alpha\beta,\alpha^\prime\beta^\prime}(\omega,\tau)$ is equivalent to Eq.~(\ref{eq:partitioned-FF}) with $T=\tau$, the distinction lies within the definition of  $G^{(i,j)}_{\alpha\beta,\alpha^\prime\beta^\prime}(\omega,\tau)$. Each of the four terms are composed of $K^2$ terms, where each constituent term captures the interactions between the $j$th and $k$th interval. More concretely, the $G^{(i,j)}_{\alpha\beta,\alpha^\prime\beta^\prime}(\omega,\tau)$ filter functions are given by
\begin{widetext}
\begin{eqnarray}
G^{(1,1)}_{\alpha\beta,\alpha^\prime\beta^\prime}(\omega,\tau) &=& \frac{\sin^2(M\omega\tau)}{\sin^2(\omega\tau)}\sum^{K}_{j,k=1} e^{i\omega [t^{(k-1)}_{2M+1} - t^{(j-1)}_{2M+1}]} F^{(1,j)}_{\alpha\beta}(-\omega,\tau)F^{(1,k)}_{\alpha^\prime\beta^\prime}(\omega,\tau)\\
G^{(1,2)}_{\alpha\beta,\alpha^\prime\beta^\prime}(\omega,\tau) &=&e^{i(M+1)\omega\tau} \frac{\sin(M\omega\tau)}{\sin(\omega\tau)}\sum^{K}_{j,k=1} e^{i\omega [t^{(k)}_{2M} - t^{(j-1)}_{2M+1}]} F^{(1,j)}_{\alpha\beta}(-\omega,\tau)F^{(2,k)}_{\alpha^\prime\beta^\prime}(-\omega,\tau)\\
G^{(2,1)}_{\alpha\beta,\alpha^\prime\beta^\prime}(\omega,\tau) &=& [G^{(1,2)}_{\alpha\beta,\alpha^\prime\beta^\prime}(\omega,\tau)]^*\\
G^{(2,2)}_{\alpha\beta,\alpha^\prime\beta^\prime}(\omega,\tau) &=&\sum^{K}_{j,k=1} e^{i\omega [t^{(k)}_{2M} - t^{(j)}_{2M}]} F^{(2,j)}_{\alpha\beta}(\omega,\tau)F^{(2,k)}_{\alpha^\prime\beta^\prime}(-\omega,\tau).
\end{eqnarray}
\end{widetext}
Each component filter function
\begin{eqnarray}
    F^{(1,j)}_{\alpha\beta}(\omega,\tau)&=&\int^{2\tau}_{0}dt\,f^{(1,j)}_{\alpha\beta}(t)e^{i\omega t}, \\
    F^{(2,j)}_{\alpha\beta}(\omega,\tau)&=&\int^{\tau}_{0}dt\,f^{(2,j)}_{\alpha\beta}(t)e^{i\omega t}.
\end{eqnarray}
denotes the Fourier Transform of the folding and post-folding switching functions, respectively. As in the global folding case, the pure folding filter function $G^{(1,1)}_{\alpha\beta,\alpha^\prime\beta^\prime}(\omega,\tau)$ is proportional to a quotient of sinusoidal functions; thus, exhibiting comb-like behavior with increasing folding repetition $M$. Despite this similarity, and many others, local folding produces a very distinct folding filter function.

In particular, local folding leads to low and high frequency localization in the filter function as $M$ grows. This behavior can be attributed to the local folding periods, with the contribution of the self-interference term $G^{(1,1)}_{\alpha\beta,\alpha^\prime\beta^\prime}(\omega,\tau)$ being most influential. This term captures two distinct types of interactions between the $j$th and $k$th folding periods that are dependent upon the modulation properties of the switching functions. Products of relatively static switching functions will lead to component filter functions that have greatest support at low frequencies, while those that demonstrate rapid fluctuations will produce filter functions that tend towards high frequencies. This effect is exacerbated by the sinusoidal (frequency comb-like) expression as the number of folding repetitions increases. 


\end{document}